\documentclass[%
reprint,
 amsmath,amssymb,
 aps,
prb,
]{revtex4-1}

\usepackage{graphicx}
\usepackage{dcolumn}
\usepackage{bm}
\usepackage{braket}

\usepackage[mathlines]{lineno}

\usepackage{todonotes}
\usepackage{lipsum}
\usepackage[english]{babel}

\begin{document}

\preprint{APS/123-QED}

\title{Spin accumulation from non-equilibrium first principles methods}

\author{Alexander Fabian}
\email{Alexander.Fabian@physik.uni-giessen.de}
\author{Michael Czerner}
\author{Christian Heiliger}
\affiliation{Institut f\"ur Theoretische Physik, Justus-Liebig-Universit\"at Giessen, Heinrich-Buff-Ring 16, 35392 Giessen, Germany}
\affiliation{Center for Materials Research (LaMa), Justus-Liebig-Universit\"at Giessen, Heinrich-Buff-Ring 16, 35392 Giessen, Germany }

\author{Hugo Rossignol}
\affiliation{HH Wills Physics Laboratory, University of Bristol, Tyndall Avenue BS8 1TL, United Kingdom}

\author{Ming-Hung Wu}
\affiliation{HH Wills Physics Laboratory, University of Bristol, Tyndall Avenue BS8 1TL, United Kingdom}
\author{Martin Gradhand}
\affiliation{HH Wills Physics Laboratory, University of Bristol, Tyndall Avenue BS8 1TL, United Kingdom}
\affiliation{Institute of Physics, Johannes Gutenberg University Mainz, 55099 Mainz, Germany}

\date{\today}

\begin{abstract}
For the technologically relevant spin Hall effect most theoretical approaches rely on the evaluation of the spin-conductivity tensor. In contrast, for most experimental configurations the generation of spin accumulation at interfaces and surfaces is the relevant quantity. Here, we directly calculate the accumulation of spins due to the spin Hall effect at the surface of a thin metallic layer, making quantitative predictions for different materials. Two distinct limits are considered, both relying on a fully relativistic Korringa-Kohn-Rostoker density functional theory method. In the semiclassical approach, we use the Boltzmann transport formalism and compare it directly to a fully quantum mechanical non-equilibrium Keldysh formalism. Restricting the calculations to the spin Hall induced, odd in spatial inversion, contribution in the limit of the relaxation time approximation we find good agreement between both methods, where deviations can be attributed to the complexity of Fermi surfaces. Finally, we compare our results to experimental values of the spin accumulation at surfaces as well as the Hall angle and find good agreement for the trend across the considered elements.

\end{abstract}

\maketitle


\section{Introduction}
\begin{figure}[tp]
	\includegraphics[width=.8\columnwidth]{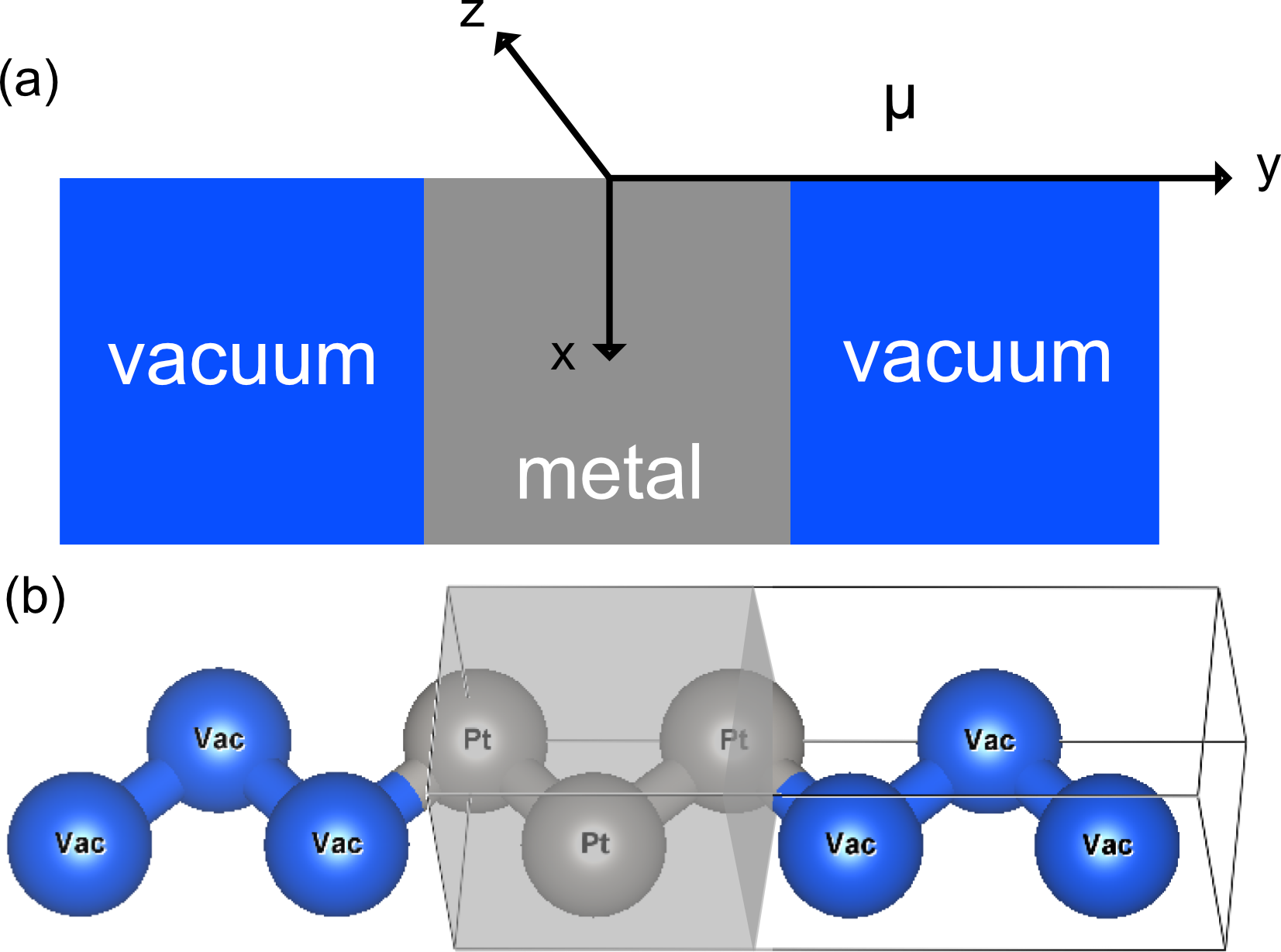}
	\caption{\label{fig:index}(a) Schematic drawing of the slab systems for the Keldysh formalism. The atomic index counts the atoms along the $y$ direction. (b)~Actual unit cell for fcc Pt. For the Keldysh formalism the box indicates the supercell.\\
	 For the Boltzmann formalism the vacuum is extended into the semi-infinite half spaces on both sides of the slab (not shown).}
\end{figure}

The spin Hall effect was first proposed in 1971 by Dyakonov and Perel.~\cite{dyakonovCurrentinducedSpinOrientation1971} Only after Hirsch~\cite{hirschSpinHallEffect1999} re-established the concept in 1999, it was experimentally observed directly in semiconductors by Kato et al.~\cite{katoObservationSpinHall2004a} and Wunderlich et al.~\cite{wunderlichExperimentalObservationSpinHall2005}. The spin Hall effect enables the generation of spin current in non-magnetic materials by passing an electric current through a system opening the route to various applications in spintronics.\cite{fukamiSpinOrbitTorque2016,pershinSpinPolarizationControl2009,hoffmannSpinHallEffects2013,liuSpinTorqueSwitchingGiant2012,kajiwaraTransmissionElectricalSignals2010,nakayamaSpinHallMagnetoresistance2013,wuHanleMagnetoresistanceRole2016,andoElectricManipulationSpin2008} Importantly, the inverse effect, generating a charge current from a spin current, or in fact a spin accumulation, gives a tool to detect spin currents electronically.~\cite{valenzuelaDirectElectronicMeasurement2006,zhaoCoherenceControlHall2006,saitohConversionSpinCurrent2006}\\
The origin of the effect is commonly divided into two contributions, the intrinsic~\cite{sinovaUniversalIntrinsicSpin2004,
guoIntrinsicSpinHall2008,
murakamiDissipationlessQuantumSpin2003,
sinovaSpinHallEffects2015} and the extrinsic mechanism. While the first derives from the intrinsic spin-orbit coupling of the pure material, the latter is mediated via spin-orbit coupling at an impurity site. For the extrinsic process, the skew or Mott scattering dominates in the dilute limit~\cite{smitSpontaneousHallEffect1955,smitSpontaneousHallEffect1958} and the side jump\cite{bergerSideJumpMechanismHall1970} scales similarly to the intrinsic mechanism with the sample resistivity.\\
Approaching the spin Hall effect theoretically is typically split into semiclassical or fully quantum mechanical approaches. In case of the semiclassical theory, the intrinsic mechanism is recast in terms of the Berry curvature~\cite{karplusHallEffectFerromagnetics1954,jungwirthAnomalousHallEffect2002,murakamiDissipationlessQuantumSpin2003} and the extrinsic, almost exclusively the skew scattering mechanism, is considered via a Boltzmann equation incorporating the vertex corrections (scattering-in term)~\cite{swihartFirstPrinciplesCalculationResidual1986,zahnImpurityScatteringQuantum2003,gradhandExtrinsicSpinHall2010}. On the other hand, the Kubo or Kubo-Streda (Kubo-Bastin) formalism has been used to consider the intrinsic mechanism~\cite{guoInitioCalculationIntrinsic2005,yaoSignChangesIntrinsic2005} or in combination with the coherent potential approximation the extrinsic mechanisms were included on equal footing.~\cite{lowitzerExtrinsicIntrinsicContributions2011} However, all approaches have in common that they almost exclusively calculate the spin Hall conductivity in a periodic crystal~\cite{guoInitioCalculationIntrinsic2009,wangInitioCalculationAnomalous2006,yaoFirstPrinciplesCalculation2004}, giving no direct access to the spin accumulation at surfaces or interfaces.
 In contrast, most experimental configurations will rely on the accumulation at interfaces and surfaces exploiting spin diffusion equations in order to extract the spin Hall conductivity.~\cite{katoObservationSpinHall2004a,stammMagnetoOpticalDetectionSpin2017,zhangSpinHallEffect2000} However, the induced spin accumulation has attracted renewed interest as the technologically relevant spin-orbit torque often relies on spin accumulation at, as well as spin currents trough, normal metal ferromagnet interfaces.~\cite{freimuthSpinorbitTorquesCo2014,wimmerFullyRelativisticDescription2016,gerantonSpinorbitTorquesSpin2016,kodderitzschLinearResponseKuboBastin2015,kosmaStrongSpinorbitTorque2020} Experimentally, it is incredibly difficult to distinguish the various contributions rendering it a challenge to optimize spin-orbit materials and the corresponding bilayer systems.~\cite{yangSpinorbitTorquePt2016,avciFieldlikeAntidampingSpinorbit2014}   \\
In this work, we directly calculate the spin accumulation induced at the surfaces of metallic thin films when a charge current is passed through the sample. We focus on the contributions with the same symmetry as the spin Hall effect namely the spin accumulation which is odd under spatial inversion~\cite{gerantonSpinorbitTorquesSpin2016,wimmerFullyRelativisticDescription2016,freimuthSpinorbitTorquesCo2014}, showing equal and opposite spin accumulations at the two surfaces of the thin metallic film. This will allow us to make contact with experimental observations and theoretical predictions of the spin Hall effect in more realistic geometries. The system is shown in Fig.~\ref{fig:index}~(a), where a charge current is driven in $z$ direction, the spin is pointing along $x$ and the accumulation is calculated in $y$ direction perpendicular to the plane of the thin film. As the atomic configuration is preserving inversion symmetry and we focus on the contributions from the clean system, it is the Fermi surface driven and odd under spatial inversions contribution~\cite{freimuthSpinorbitTorquesCo2014,wimmerFullyRelativisticDescription2016}, which is linear in the applied longitudinal current for which we make quantitative predictions in a series of metallic systems.
On one hand, we go beyond the semiclassical approach~\cite{gerantonSpinorbitTorquesSpin2016} previously applied to bi-layer systems using a fully quantum mechanical Keldysh formalism based on non-equilibrium Green's functions. On the other hand, we apply this formalism to real materials in a fully ab initio density functional (DFT) frame work going a step further than earlier work of the spin accumulation in non-equilibrium description which were restricted to a model Hamiltonian.~\cite{nikolicNonequilibriumSpinHall2005,nikolicImagingMesoscopicSpin2006} To validate our method, we compare it to a semiclassical approach relying on the Boltzmann formalism. 

After a brief introduction of both methods we will present exemplary results and compare the induced spin accumulation to experimental findings. Furthermore, we will analyse the common trends across the elements with respect to the charge to spin current conversion efficiency.

\section{Theory}
The electronic structure is calculated via a fully relativistic Korringa-Kohn-Rostoker (KKR) density functional theory method~\cite{zabloudilElectronScatteringSolid2005}. Both band structure methods, for the semiclassical approach~\cite{gradhandSpinPolarizationFermi2009, gradhandFullyRelativisticInitio2010} and the Keldysh formalism~\cite{heiligerImplementationNonequilibriumGreen2008,
franzImplementationNonequilibriumVertex2013,
mahrImplementationMethodCalculating2017}, have been introduced earlier. Here, we only highlight the adjustments and relevant expressions used to express the steady-state magnetization density. 

\subsection{Keldysh formalism}
For the Keldysh formalism the system is divided into three parts, Left (L), Center (C) and Right (R) region. The left and right parts work as semi infinite leads. The leads are considered to be in an equilibrium  state. Their influence to the center region is accounted for by the corresponding self energies $\Sigma_{L/R}$. The Fermi levels $E_F$ are the same, if they are of the same material. When applying a bias voltage, the levels of the chemical potential change to $\eta_{>/<}=E_F\pm\frac{e\Delta\varphi}{2}$ accordingly. 
In the range of $[\eta_<,\eta_>]$ the fully relativistic electron density and magnetization density is calculated as\cite{zabloudilElectronScatteringSolid2005} 
\begin{align}
\rho(\vec r)&=\frac{1}{2\pi}\int_{\eta_<}^{\eta_>} \Bra{\vec r}\mathrm{Tr}\left[ G(E)\Gamma(E) G^\dagger(E)\right]\Ket{\vec r} \; \mathrm d E\ \text{,}\\
m^{(i)}(\vec r)&=\frac{\mu_B}{2\pi}\int_{\eta_<}^{\eta_>} \Bra{\vec r}\mathrm{Tr}\left[ \beta\varsigma_i G(E)\Gamma(E) G^\dagger(E)\right]\Ket{\vec r} \; \mathrm d E\ \text{,}
\end{align} 
respectively. Here, $G(E)$ is the Green's function of the center area, and $\Gamma=i(\Sigma(E)-\Sigma^\dagger(E))$ the broadening function, where $\Sigma(E)=\Sigma_L(E)+\Sigma_R(E)$,  
\[
\beta=\begin{pmatrix}
	I_2 & 0 \\ 0 & -I_2
\end{pmatrix}, \quad
\varsigma_i =\begin{pmatrix}
	\sigma_i & 0 \\ 0 & \sigma_i
\end{pmatrix},
\] 
$I_2$ is the $2\times2$ unity matrix, and $\sigma_i$ are the Pauli spin matrices with $i \in \{x,y,z\}$. In the so-called one-shot calculations only the magnetization at the Fermi level is considered for vanishing bias voltage, that is
\[ m_i(\vec r) = \frac{\mu_B}{2\pi} \mathrm{Tr}[ \beta \varsigma_i G(E_\text{F})\Gamma(E_\text{F}) G^\dagger(E_\text{F})] e \Delta \varphi \ \text{.} \]
Finally, the magnetic moment due to spin accumulation $a_x(\mu)$ is evaluated by integrating $m_x(\vec r)$ over the volume $V_\mu$ of the atomic sphere at atomic index $\mu$:  
\begin{equation}
a_x(\mu)=\int_{V_\mu}m_x(\vec r) \; \mathrm d V.
\end{equation} 
The current density is calculated via the Landauer-B\"uttiker formula in the case of a vanishing bias voltage\cite{dattaElectronicTransportMesoscopic1995}
\begin{equation}
j_z=\frac{e^2}{A \hbar}T(E_\text{F})\Delta \varphi
\end{equation}
assuming the transmission $T(E)=\mathrm{Tr} [\Gamma_LG\Gamma_RG^\dagger]$ is nearly constant in the range of $\Delta E = e \Delta \varphi$. Here, $A$ is the area of the super cell in $x$ and $y$ direction.
 
\subsection{Boltzmann formalism}

Within the Boltzmann formalism the spin accumulation is expressed as Fermi surface integral~\cite{gerantonSpinorbitTorquesSpin2016}. For 2D systems the spin accumulation is expressed as~\cite{herschbachEnhancementSpinHall2012}
\begin{equation}
\vec a=\underline \chi_\mu \cdot \vec E=-\frac{e \mu_\text{B}}{\hbar}\frac{V}{d(2\pi)^2} \int_{E_\text{F}} \frac{\mathrm{d}l } { | \vec{v}_{\vec{k}} | }\left(\vec s_{\vec{k}}(\mu) \circ \tau_{\vec k} \vec{v}_{\vec k}\right)\cdot \vec E\ \text{,}
\end{equation}
where $V$ is the volume of the cell, $d$ the thickness of the film, $v_{\vec{k}}$ the group velocity at $\vec{k}$, $\vec s_{\vec k}$ the expectation value of the spin operator, and $\vec E$ the applied electric field. Because of degenerate states, the spin operator exhibits off-diagonal elements. A gauge transformation is applied, such that these off-diagonal elements vanish.
The current density is given by
\begin{equation}
\vec j = \underline \sigma\cdot \vec E=-\frac{e^2 }{\hbar}\frac{1}{d(2\pi)^2}  \int_{E_\text{F}} \frac{\mathrm{d}l} { | \vec{v}_{\vec{k}} | } \left(\vec v_{\vec{k}} \circ \tau_{\vec k} \vec{v}_{\vec k}\right) \cdot \vec E\ \text{.}
\end{equation}
Importantly both scale linearly with the relaxation time. In the chosen geometry $\vec{j} = j\vec{e}_z$ and $\vec{E}=E_z \vec{e}_z$, and by using the relaxation time approximation $\tau _{\vec{k}} = \tau $, the relevant expressions can be simplified as 
\begin{equation}
\label{eq:aboltzsimp}
a_x(\mu)=\chi_{xz}E_z=\frac{e}{\hbar}\frac{\mu_\text{B}  V\tau E_z}{d(2\pi)^2} \int_{E_\text{F}} \frac{\mathrm{d}l } { | \vec{v}_{\vec{k}} | } s_{x,\vec k}(\mu)v_{z,\vec k}
\end{equation}
and 
\begin{equation}
\label{eq:jboltzsimp}
j_z=\frac{e^2}{\hbar} \frac{ \tau E_z}{d(2\pi)^2}\int_{E_\text{F}} \frac{v_{z,\vec k} v_{z,\vec k}} { | \vec{v}_{\vec{k}} | }\mathrm{d}l=\frac{e^2}{\hbar} \frac{ \tau E_z}{d(2\pi)^2} \langle v_z^2 \rangle.
\end{equation}
This manoeuvre will allow us to remove the direct dependence of the spin accumulation on the relaxation time $\tau E_z$ replacing it with the current density
\begin{equation}
\label{eq:aboltzsimp2}
\frac{a_x(\mu)}{\mu_\text{B}}=\frac{j_z}{e}\frac{V}{\langle v_z^2\rangle} \int_{E_\text{F}} \frac{\mathrm{d}l } { | \vec{v}_{\vec{k}} | } s_{x,\vec k}(\mu)v_{z,\vec k}\ \text{.}
\end{equation}
 Thus the spin accumulation will scale linearly with the current density which in turn can be calculated within the Keldysh formalism. This will allow for direct mapping between the two methods.

\subsection{Computational details}

Slight differences in the two implementations lead to differences in the geometrical construction. For the Keldysh formalism the starting point are self-consistently calculated equilibrium potentials, which are obtained in a super cell approach including atomic spheres and vacuum spheres to form the thin film geometry. For the transport calculations, the super cell is connected to semi-infinite leads from the left and right side along the transport direction ($z$-direction). 
 The corresponding cells are schematically shown in Fig.~\ref{fig:index}. In the following, a one-step non-equilibrium Keldysh formalism at the Fermi energy is used to find the steady-state densities from these potentials. The applied voltage is chosen reasonably small at $\Delta \varphi=10^{-4}~\textrm{Ry}/e$, in order to agree with the approximation of vanishing applied electric field in the linear response regime as assumed in the Boltzmann approach. The transmission function as well as the magnetization density does not change significantly in the small bias window.

For the Boltzmann formalism the construction is based on a slab calculation with semi-infinite vacuum attached perpendicular to the film. After obtaining the self-consistent potentials, the Fermi surface parameters such as the $\vec{k}$-resolved band velocities and spin expectation values are calculated to find the spin accumulation according to Eq.~(\ref{eq:aboltzsimp2}). Given the linear scaling of the spin accumulation with the current density in the Boltzmann formalism we insert the current density found within the Keldysh approach to facilitate direct comparison. 
 
As a note of caution we would like to highlight the differences between the two approaches with respect to the origin of charge resistivities. Within the Landauer-B\"uttiker approach the finite conductance stems from a contact resistance at the interfaces of the leads. This contact resistance is also often referred to as Sharvin resistance~\cite{sharvinPossibleMethodStudying1965}. Naturally, it does not depend on the length of the transport system but only on the number of available transport channels.
 In contrast, for the Boltzmann approach the contact resistance is ignored and the whole resistance originates from scattering in the volume. In our comparison we adjust $j$ such that it fits the Sharvin resistance of the Landauer-B\"uttiker approach. As such the mechanism for the finite currents is different in both approaches, however the resulting current density itself is the same, driving the spin accumulation at the surfaces.

Here, we considered cubic systems (fcc and bcc). As we apply a bias in $z$ direction, the only relevant element of the spin accumulation is $a_x(\mu)$ and for convenience we are going to omit the index $x$ in the following. The axes of the coordinate systems are aligned parallel to the $\langle 100\rangle$ axes of the crystals.   
\section{Results and Discussion}

\begin{figure*}[htp]
	\includegraphics[width=\columnwidth]{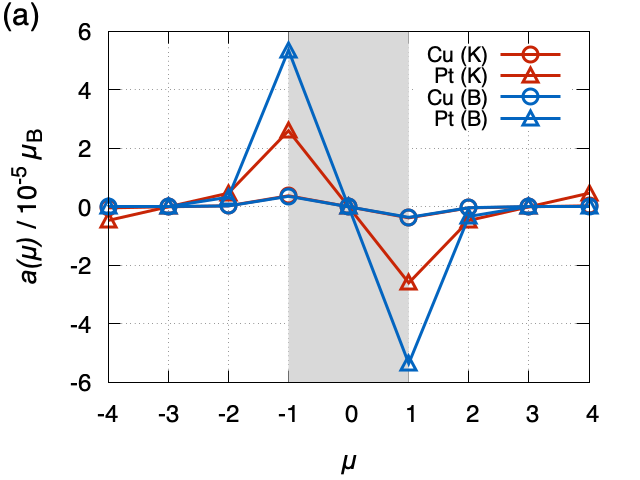}
	\includegraphics[width=\columnwidth]{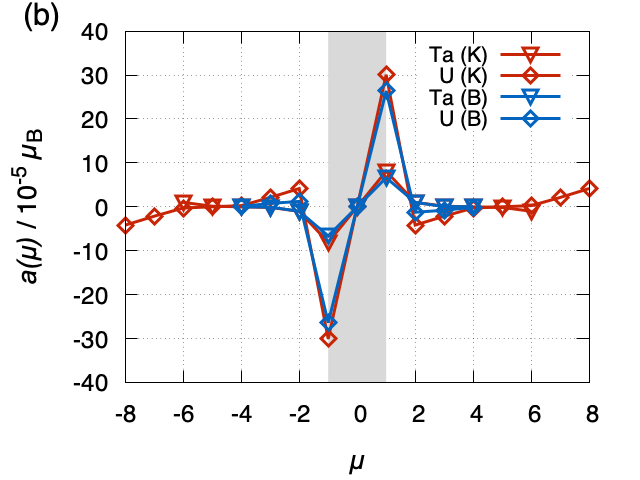}
	\caption{\label{fig:fccbcc} Plot of the magnetic moment per atom for representative (a) fcc systems and (b) bcc systems. Blue refers to Keldysh (K) calculated values, red to Boltzmann (B) calculated values. The thin film is highlighted in grey. Each line shows the same antisymmetric behaviour. Note, that in (a) the Keldysh and Boltzmann values for Cu overlap. }
\end{figure*}

The resulting spin accumulation $a(\mu)$ as a function of the atomic position index $\mu$  is exemplary shown in Fig.~\ref{fig:fccbcc}
 for the (a) fcc (Cu, Pt) and (b) bcc (Ta and U) systems comparing the Keldysh (K) and Boltzmann (B) formalism, respectively. The position index is chosen such that the central atom of the film is labelled as 0. 

The general behaviour of the accumulation $a(\mu)$ in Fig.~\ref{fig:fccbcc} is the same for all considered elements as well as between the two methods. This is largely enforced by symmetry since atoms $\mu=\pm 1$ have equal and opposite spin accumulation leading to vanishing magnetization for the central atom. For easier comparison we summarize the maximum spin accumulation $a(\mu=-1)$ for the various systems as well as the two methods in Table~\ref{tab:accu}. As expected, the spin accumulation increases with increasing atomic weight corresponding to enhanced spin-orbit coupling. While this is true in general with U showing the largest effect it is not correct in the details. The spin accumulation for Ag is smaller than for Cu and for Ta we find a surprisingly large spin accumulation. Such details would be difficult to predict from simplified models. Comparing the Boltzmann to the Keldysh formalism the agreement is perfect for the noble metals, with their simple Fermi surfaces, but starts to deviate for the more complex systems Ta, Pd, Pt, and U. Nevertheless, the sign as well as the overall magnitude is still in remarkable agreement.

\begin{table*}
	\caption{\label{tab:accu} First Extrema of the  spin accumulation calculated by Boltzmann and Keldysh formalisms as well as the Keldysh current density for the small systems.  Comparison between $a/j$ and spin Hall angle $\theta_\text{SH}^\text{exp}$\cite{wangScalingSpinHall2014}. Intrinsic spin Hall conductivities from calculations are shown for reference. }	

\begin{ruledtabular}
\begin{tabular}{cddddddd}
	element & \multicolumn{2}{c}{ $a(\mu=-1)\; [10^{-6}\mu_B]$} & \multicolumn{1}{c}{$j$ $[10^{12} \text{A}\text{m}^{-2}]$} & \multicolumn{2}{c}{ $a/j \; [10^{-17}\mu_B \text{m}^2/\text{A}]$} & \multicolumn{1}{c}{$\theta^\text{exp}_\text{SH}$\cite{wangScalingSpinHall2014}} & \multicolumn{1}{c}{$\sigma^\text{theo}_{\text{SH}}$} \\
	& \multicolumn{1}{c}{Boltzmann} & \multicolumn{1}{c}{Keldysh} & & \multicolumn{1}{c}{Boltzmann} & \multicolumn{1}{c}{Keldysh} & \multicolumn{1}{c}{$[\%]$} &   \multicolumn{1}{c}{$[(\hbar / e)~\Omega^{-1} \text{cm}^{-1}]$} \\ \colrule
	Cu (fcc)& 3.60 & 3.77 & 1.50        &    0.24 & 0.25 & 0.32 &  - \\    
	Ag (fcc)& 2.65 & 2.41 & 1.16        &    0.23 & 0.21 & 0.68 & - \\    
	Au (fcc)& 17.62 & 16.33 & 1.33      &    1.32 & 1.23 & 8.4 & 400\cite{guoInitioCalculationIntrinsic2009} \\    
	Ta (bcc) & -66.19 & -80.99 & 1.17   &     -5.66 & -6.92 & -7.1  & -142\cite{qiaoCalculationIntrinsicSpin2018} \\    
	Pd (fcc)& 31.33 & 10.10 & 1.12      &    2.80 & 0.90 & -  & 1400\cite{guoInitioCalculationIntrinsic2009} \\    
	Pt (fcc)& 53.53 & 25.99 & 1.79      &    2.99 & 1.45 & 10.00 & 2000\cite{guoIntrinsicSpinHall2008}  \\    
	U (bcc)& -263.9 & -300.9  & 1.93  &     -13.7 & -15.6  & - & -402\cite{wuSpindependentTransportUranium2020}     \\
\end{tabular}
\end{ruledtabular}
\end{table*}

We believe this correlation between Fermi surface complexity (see Fig.~5 in the Supplemental Material\footnote{Supplemental Material}) and agreement between the two methods not to be a simple numerical artefact. In the Keldysh formalism we only consider the ballistic transport where each band contributes equally to the electronic transport. In contrast, the Boltzmann formalism relies on electron scattering and the weighting in any Fermi surface integral will depend on the $\vec k$-dependent band velocity in transport direction. For more complex structures the variations of the absolute value of the band velocity on the Fermi surface are much more pronounced (Ta, U, Pd, Pt) than for the simple metals Au, Ag, and Cu (see Fig.~5 in Supplemental Material~\cite{Note1}). As the Keldysh formalism considers the ballistic limit, entirely ignoring any scattering, the results can be interpreted as the clean limit. For elements with simple Fermi surfaces and subsequently least changing Fermi velocity, the results obtained within the Boltzmann formalism nevertheless match well. Considering they are computationally much faster they can serve as a quick alternative to the cumbersome full non-equilibrium Keldysh formalism.

\begin{figure}[bp]
	\includegraphics[width=.5\textwidth]{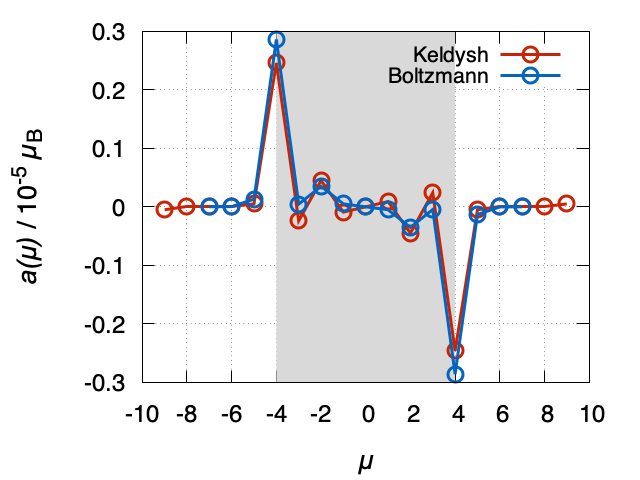}
	\caption{\label{fig:Cu_big} Magnetization for a thin film of 9 Cu atoms (highlighted in grey). While the overall behaviour is the same in both methods there are differences in the absolute values. With the Boltzmann method one gets higher values on the first peak but lower values on the second peak compared to the Keldysh method. }
\end{figure}

So far we considered rather thin layers with limited access to the decay length of the spin accumulation within the thin film. To investigate this point further we consider three larger systems, Cu, Pt, and U, with nine layers of atoms (c.f. Fig. \ref{fig:Cu_big} and Supplemental Material\cite{Note1}). 
For Cu the decay of the spin accumulation is remarkably strong, happening within 3 layers and is in excellent agreement between the two methods. In contrast, the decay appears much slower for Pt and even more so for U (see Fig.~3 (c) in Supplemental Material\cite{Note1} ) again consistent between the two methods.
	 
In order to validate our results we compare to recent experiments, where the spin accumulation of Pt thin films was directly measured by MOKE~\cite{stammMagnetoOpticalDetectionSpin2017}. In that experiment a strong thickness dependence was established with a value of $a/j=5\times 10^{-16}~\mu_B\text{A}^{-1} \text{m}^2 $ for samples with a thickness $t>40~\text{nm}$. Extrapolating the experimental data (Eq. 1 in Ref.~\onlinecite{stammMagnetoOpticalDetectionSpin2017}) to the film thickness of $t=0.39~\text{nm}$ considered here, yields a result of $a/j= 1.05 \times 10^{-17}~\mu_B \text{A}^{-1} \text{m}^2$ in rather good agreement to our result $a/j = 1.45\times 10^{-17}~\mu_B \text{A}^{-1} \text{m}^2 $.
While measurements of spin Hall angles and spin Hall conductivities are widely available, to our knowledge, such direct numerical measurements of the spin accumulation for other systems are very sparse. It is therefore difficult to compare the results from our methods directly to literature values. It appears natural to compare to spin Hall conductivities or spin Hall angles predicted theoretically or measured experimentally. However, this holds multiple caveats. For example, theoretically predicted intrinsic conductivities are bulk calculations ignoring the fact that any spin accumulation will depend on the actual surface geometry and film thickness. While sign changes and over all magnitudes ought to be in agreement significant variations are possible in the details. As summarized in Table~\ref{tab:accu} the signs are in agreement between the spin accumulations and the intrinsic spin Hall conductivities but the high spin accumulation for Ta and U cannot trivially be predicted from the conductivities. On the other hand experimental results for the spin Hall angles tend to vary significantly over the various experimental techniques and sample preparations, which will often involve varying degrees of extrinsic mechanisms contributing to the overall effect.\cite{sagastaTuningSpinHall2016} Consequently any comparison should focus on one technique with similar sample preparation only.
Choosing a spin pumping experiment, in which most of the considered metals were investigated under similar conditions \cite{wangScalingSpinHall2014} the trend for $a/j$ and $\theta_\text{SH}^\text{exp}$ in Table~\ref{tab:accu} is quite consistent for
systems with simpler Fermi surfaces (Cu, Ag, Au). Similarly, 
Ta, Au, and Pt show increasing spin Hall angles in the same order of magnitude, with a sign change occurring for Ta. 

\section{Conclusion}
We extended existing theoretical frameworks to capture the spin Hall effect induced spin accumulation in various metallic thin films via a fully non-equilibrium Keldysh formalism. We tested this new approach against a linearized Boltzmann approach as well as experimental findings and found remarkable agreement in all cases, reproducing all sign changes and predicting the same trends. Where the two theoretical approaches differ most is the atom resolved spin accumulation in thicker films especially for systems with complex Fermi surfaces whereas for Cu we find an excellent agreement. This new methodology will enable us to make more direct contact with experiments where instead of the conductivities derived from periodic crystals it is the spin accumulation at interfaces and surfaces as well as the spin current through interfaces which are the relevant driving mechanisms of for example magnetisation reversal in ferromagnets. In this first and most important step we have established that the developed methodology reproduces the spin Hall induced spin accumulation in the thin metallic films well across different frameworks and in comparison to experiment. This will open up broad opportunities to explore the effect in more complex interfaces as well as under the influence of impurities making even more direct contact with experimental realities. Incorporating, inversion asymmetry and contributions even under spatial inversion symmetry~\cite{freimuthSpinorbitTorquesCo2014,wimmerFullyRelativisticDescription2016} will give access to spin galvanic effects~\cite{skinnerComplementarySpinHallInverse2015} while investigating the additional influence of impurities and the additional Mott scattering~\cite{kodderitzschLinearResponseKuboBastin2015}. In all those cases, the full non-equilibrium description adds additional complexity with the possibility of finite bias across the sample geometry.  

\section{Acknowledgements}
A. F., M. C. and C. H. acknowledge computational resources provided by the HPC Core Facility and the HRZ of the Justus-Liebig-University Giessen.
Further, they would like to thank Marcel Giar and Philipp Risius of HPC-Hessen, funded by the State Ministry of Higher Education, Research and the Arts, for technical support.
M.-H. W., H. R. and M. G. carried out their computational work using the computational facilities of the Advanced Computing Research Centre, University of Bristol.
M.G. thanks the visiting professorship program of the Centre for Dynamics and Topology at Johannes Gutenberg-University Mainz.


\bibliography{bibliography}

\begin{thebibliography}{59}%
\makeatletter
\providecommand \@ifxundefined [1]{%
 \@ifx{#1\undefined}
}%
\providecommand \@ifnum [1]{%
 \ifnum #1\expandafter \@firstoftwo
 \else \expandafter \@secondoftwo
 \fi
}%
\providecommand \@ifx [1]{%
 \ifx #1\expandafter \@firstoftwo
 \else \expandafter \@secondoftwo
 \fi
}%
\providecommand \natexlab [1]{#1}%
\providecommand \enquote  [1]{``#1''}%
\providecommand \bibnamefont  [1]{#1}%
\providecommand \bibfnamefont [1]{#1}%
\providecommand \citenamefont [1]{#1}%
\providecommand \href@noop [0]{\@secondoftwo}%
\providecommand \href [0]{\begingroup \@sanitize@url \@href}%
\providecommand \@href[1]{\@@startlink{#1}\@@href}%
\providecommand \@@href[1]{\endgroup#1\@@endlink}%
\providecommand \@sanitize@url [0]{\catcode `\\12\catcode `\$12\catcode
  `\&12\catcode `\#12\catcode `\^12\catcode `\_12\catcode `\%12\relax}%
\providecommand \@@startlink[1]{}%
\providecommand \@@endlink[0]{}%
\providecommand \url  [0]{\begingroup\@sanitize@url \@url }%
\providecommand \@url [1]{\endgroup\@href {#1}{\urlprefix }}%
\providecommand \urlprefix  [0]{URL }%
\providecommand \Eprint [0]{\href }%
\providecommand \doibase [0]{http://dx.doi.org/}%
\providecommand \selectlanguage [0]{\@gobble}%
\providecommand \bibinfo  [0]{\@secondoftwo}%
\providecommand \bibfield  [0]{\@secondoftwo}%
\providecommand \translation [1]{[#1]}%
\providecommand \BibitemOpen [0]{}%
\providecommand \bibitemStop [0]{}%
\providecommand \bibitemNoStop [0]{.\EOS\space}%
\providecommand \EOS [0]{\spacefactor3000\relax}%
\providecommand \BibitemShut  [1]{\csname bibitem#1\endcsname}%
\let\auto@bib@innerbib\@empty
\bibitem [{\citenamefont {Dyakonov}\ and\ \citenamefont
  {Perel}(1971)}]{dyakonovCurrentinducedSpinOrientation1971}%
  \BibitemOpen
  \bibfield  {author} {\bibinfo {author} {\bibfnamefont {M.~I.}\ \bibnamefont
  {Dyakonov}}\ and\ \bibinfo {author} {\bibfnamefont {V.~I.}\ \bibnamefont
  {Perel}},\ }\href {\doibase 10.1016/0375-9601(71)90196-4} {\bibfield
  {journal} {\bibinfo  {journal} {Phys. Lett. A}\ }\textbf {\bibinfo {volume}
  {35}},\ \bibinfo {pages} {459} (\bibinfo {year} {1971})}\BibitemShut
  {NoStop}%
\bibitem [{\citenamefont {Hirsch}(1999)}]{hirschSpinHallEffect1999}%
  \BibitemOpen
  \bibfield  {author} {\bibinfo {author} {\bibfnamefont {J.~E.}\ \bibnamefont
  {Hirsch}},\ }\href {\doibase 10.1103/PhysRevLett.83.1834} {\bibfield
  {journal} {\bibinfo  {journal} {Phys. Rev. B}\ }\textbf {\bibinfo {volume}
  {83}},\ \bibinfo {pages} {1834} (\bibinfo {year} {1999})}\BibitemShut
  {NoStop}%
\bibitem [{\citenamefont {Kato}\ \emph {et~al.}(2004)\citenamefont {Kato},
  \citenamefont {Myers}, \citenamefont {Gossard},\ and\ \citenamefont
  {Awschalom}}]{katoObservationSpinHall2004a}%
  \BibitemOpen
  \bibfield  {author} {\bibinfo {author} {\bibfnamefont {Y.~K.}\ \bibnamefont
  {Kato}}, \bibinfo {author} {\bibfnamefont {R.~C.}\ \bibnamefont {Myers}},
  \bibinfo {author} {\bibfnamefont {A.~C.}\ \bibnamefont {Gossard}}, \ and\
  \bibinfo {author} {\bibfnamefont {D.~D.}\ \bibnamefont {Awschalom}},\ }\href
  {\doibase 10.1126/science.1105514} {\bibfield  {journal} {\bibinfo  {journal}
  {Science}\ }\textbf {\bibinfo {volume} {306}},\ \bibinfo {pages} {1910}
  (\bibinfo {year} {2004})}\BibitemShut {NoStop}%
\bibitem [{\citenamefont {Wunderlich}\ \emph {et~al.}(2005)\citenamefont
  {Wunderlich}, \citenamefont {Kaestner}, \citenamefont {Sinova},\ and\
  \citenamefont {Jungwirth}}]{wunderlichExperimentalObservationSpinHall2005}%
  \BibitemOpen
  \bibfield  {author} {\bibinfo {author} {\bibfnamefont {J.}~\bibnamefont
  {Wunderlich}}, \bibinfo {author} {\bibfnamefont {B.}~\bibnamefont
  {Kaestner}}, \bibinfo {author} {\bibfnamefont {J.}~\bibnamefont {Sinova}}, \
  and\ \bibinfo {author} {\bibfnamefont {T.}~\bibnamefont {Jungwirth}},\ }\href
  {\doibase 10.1103/PhysRevLett.94.047204} {\bibfield  {journal} {\bibinfo
  {journal} {Phys. Rev. Lett.}\ }\textbf {\bibinfo {volume} {94}},\ \bibinfo
  {pages} {047204} (\bibinfo {year} {2005})}\BibitemShut {NoStop}%
\bibitem [{\citenamefont {Fukami}\ \emph {et~al.}(2016)\citenamefont {Fukami},
  \citenamefont {Anekawa}, \citenamefont {Zhang},\ and\ \citenamefont
  {Ohno}}]{fukamiSpinOrbitTorque2016}%
  \BibitemOpen
  \bibfield  {author} {\bibinfo {author} {\bibfnamefont {S.}~\bibnamefont
  {Fukami}}, \bibinfo {author} {\bibfnamefont {T.}~\bibnamefont {Anekawa}},
  \bibinfo {author} {\bibfnamefont {C.}~\bibnamefont {Zhang}}, \ and\ \bibinfo
  {author} {\bibfnamefont {H.}~\bibnamefont {Ohno}},\ }\href {\doibase
  10.1038/nnano.2016.29} {\bibfield  {journal} {\bibinfo  {journal} {Nat.
  Nanotechnol.}\ }\textbf {\bibinfo {volume} {11}},\ \bibinfo {pages} {621}
  (\bibinfo {year} {2016})}\BibitemShut {NoStop}%
\bibitem [{\citenamefont {Pershin}\ \emph {et~al.}(2009)\citenamefont
  {Pershin}, \citenamefont {Sinitsyn}, \citenamefont {Kogan}, \citenamefont
  {Saxena},\ and\ \citenamefont {Smith}}]{pershinSpinPolarizationControl2009}%
  \BibitemOpen
  \bibfield  {author} {\bibinfo {author} {\bibfnamefont {Y.~V.}\ \bibnamefont
  {Pershin}}, \bibinfo {author} {\bibfnamefont {N.~A.}\ \bibnamefont
  {Sinitsyn}}, \bibinfo {author} {\bibfnamefont {A.}~\bibnamefont {Kogan}},
  \bibinfo {author} {\bibfnamefont {A.}~\bibnamefont {Saxena}}, \ and\ \bibinfo
  {author} {\bibfnamefont {D.~L.}\ \bibnamefont {Smith}},\ }\href {\doibase
  10.1063/1.3180494} {\bibfield  {journal} {\bibinfo  {journal} {Appl. Phys.
  Lett.}\ }\textbf {\bibinfo {volume} {95}},\ \bibinfo {pages} {022114}
  (\bibinfo {year} {2009})}\BibitemShut {NoStop}%
\bibitem [{\citenamefont {Hoffmann}(2013)}]{hoffmannSpinHallEffects2013}%
  \BibitemOpen
  \bibfield  {author} {\bibinfo {author} {\bibfnamefont {A.}~\bibnamefont
  {Hoffmann}},\ }\href {\doibase 10.1109/TMAG.2013.2262947} {\bibfield
  {journal} {\bibinfo  {journal} {IEEE Trans. Magn.}\ }\textbf {\bibinfo
  {volume} {49}},\ \bibinfo {pages} {5172} (\bibinfo {year}
  {2013})}\BibitemShut {NoStop}%
\bibitem [{\citenamefont {Liu}\ \emph {et~al.}(2012)\citenamefont {Liu},
  \citenamefont {Pai}, \citenamefont {Li}, \citenamefont {Tseng}, \citenamefont
  {Ralph},\ and\ \citenamefont {Buhrman}}]{liuSpinTorqueSwitchingGiant2012}%
  \BibitemOpen
  \bibfield  {author} {\bibinfo {author} {\bibfnamefont {L.}~\bibnamefont
  {Liu}}, \bibinfo {author} {\bibfnamefont {C.-F.}\ \bibnamefont {Pai}},
  \bibinfo {author} {\bibfnamefont {Y.}~\bibnamefont {Li}}, \bibinfo {author}
  {\bibfnamefont {H.~W.}\ \bibnamefont {Tseng}}, \bibinfo {author}
  {\bibfnamefont {D.~C.}\ \bibnamefont {Ralph}}, \ and\ \bibinfo {author}
  {\bibfnamefont {R.~A.}\ \bibnamefont {Buhrman}},\ }\href {\doibase
  10.1126/science.1218197} {\bibfield  {journal} {\bibinfo  {journal}
  {Science}\ }\textbf {\bibinfo {volume} {336}},\ \bibinfo {pages} {555}
  (\bibinfo {year} {2012})}\BibitemShut {NoStop}%
\bibitem [{\citenamefont {Kajiwara}\ \emph {et~al.}(2010)\citenamefont
  {Kajiwara}, \citenamefont {Harii}, \citenamefont {Takahashi}, \citenamefont
  {Ohe}, \citenamefont {Uchida}, \citenamefont {Mizuguchi}, \citenamefont
  {Umezawa}, \citenamefont {Kawai}, \citenamefont {Ando}, \citenamefont
  {Takanashi}, \citenamefont {Maekawa},\ and\ \citenamefont
  {Saitoh}}]{kajiwaraTransmissionElectricalSignals2010}%
  \BibitemOpen
  \bibfield  {author} {\bibinfo {author} {\bibfnamefont {Y.}~\bibnamefont
  {Kajiwara}}, \bibinfo {author} {\bibfnamefont {K.}~\bibnamefont {Harii}},
  \bibinfo {author} {\bibfnamefont {S.}~\bibnamefont {Takahashi}}, \bibinfo
  {author} {\bibfnamefont {J.}~\bibnamefont {Ohe}}, \bibinfo {author}
  {\bibfnamefont {K.}~\bibnamefont {Uchida}}, \bibinfo {author} {\bibfnamefont
  {M.}~\bibnamefont {Mizuguchi}}, \bibinfo {author} {\bibfnamefont
  {H.}~\bibnamefont {Umezawa}}, \bibinfo {author} {\bibfnamefont
  {H.}~\bibnamefont {Kawai}}, \bibinfo {author} {\bibfnamefont
  {K.}~\bibnamefont {Ando}}, \bibinfo {author} {\bibfnamefont {K.}~\bibnamefont
  {Takanashi}}, \bibinfo {author} {\bibfnamefont {S.}~\bibnamefont {Maekawa}},
  \ and\ \bibinfo {author} {\bibfnamefont {E.}~\bibnamefont {Saitoh}},\ }\href
  {\doibase 10.1038/nature08876} {\bibfield  {journal} {\bibinfo  {journal}
  {Nature}\ }\textbf {\bibinfo {volume} {464}},\ \bibinfo {pages} {262}
  (\bibinfo {year} {2010})}\BibitemShut {NoStop}%
\bibitem [{\citenamefont {Nakayama}\ \emph {et~al.}(2013)\citenamefont
  {Nakayama}, \citenamefont {Althammer}, \citenamefont {Chen}, \citenamefont
  {Uchida}, \citenamefont {Kajiwara}, \citenamefont {Kikuchi}, \citenamefont
  {Ohtani}, \citenamefont {Gepr{\"a}gs}, \citenamefont {Opel}, \citenamefont
  {Takahashi}, \citenamefont {Gross}, \citenamefont {Bauer}, \citenamefont
  {Goennenwein},\ and\ \citenamefont
  {Saitoh}}]{nakayamaSpinHallMagnetoresistance2013}%
  \BibitemOpen
  \bibfield  {author} {\bibinfo {author} {\bibfnamefont {H.}~\bibnamefont
  {Nakayama}}, \bibinfo {author} {\bibfnamefont {M.}~\bibnamefont {Althammer}},
  \bibinfo {author} {\bibfnamefont {Y.-T.}\ \bibnamefont {Chen}}, \bibinfo
  {author} {\bibfnamefont {K.}~\bibnamefont {Uchida}}, \bibinfo {author}
  {\bibfnamefont {Y.}~\bibnamefont {Kajiwara}}, \bibinfo {author}
  {\bibfnamefont {D.}~\bibnamefont {Kikuchi}}, \bibinfo {author} {\bibfnamefont
  {T.}~\bibnamefont {Ohtani}}, \bibinfo {author} {\bibfnamefont
  {S.}~\bibnamefont {Gepr{\"a}gs}}, \bibinfo {author} {\bibfnamefont
  {M.}~\bibnamefont {Opel}}, \bibinfo {author} {\bibfnamefont {S.}~\bibnamefont
  {Takahashi}}, \bibinfo {author} {\bibfnamefont {R.}~\bibnamefont {Gross}},
  \bibinfo {author} {\bibfnamefont {G.~E.~W.}\ \bibnamefont {Bauer}}, \bibinfo
  {author} {\bibfnamefont {S.~T.~B.}\ \bibnamefont {Goennenwein}}, \ and\
  \bibinfo {author} {\bibfnamefont {E.}~\bibnamefont {Saitoh}},\ }\href
  {\doibase 10.1103/PhysRevLett.110.206601} {\bibfield  {journal} {\bibinfo
  {journal} {Phys. Rev. Lett.}\ }\textbf {\bibinfo {volume} {110}},\ \bibinfo
  {pages} {206601} (\bibinfo {year} {2013})}\BibitemShut {NoStop}%
\bibitem [{\citenamefont {Wu}\ \emph {et~al.}(2016)\citenamefont {Wu},
  \citenamefont {Zhang}, \citenamefont {Wan}, \citenamefont {Tao},
  \citenamefont {Huang}, \citenamefont {Kong},\ and\ \citenamefont
  {Han}}]{wuHanleMagnetoresistanceRole2016}%
  \BibitemOpen
  \bibfield  {author} {\bibinfo {author} {\bibfnamefont {H.}~\bibnamefont
  {Wu}}, \bibinfo {author} {\bibfnamefont {X.}~\bibnamefont {Zhang}}, \bibinfo
  {author} {\bibfnamefont {C.~H.}\ \bibnamefont {Wan}}, \bibinfo {author}
  {\bibfnamefont {B.~S.}\ \bibnamefont {Tao}}, \bibinfo {author} {\bibfnamefont
  {L.}~\bibnamefont {Huang}}, \bibinfo {author} {\bibfnamefont {W.~J.}\
  \bibnamefont {Kong}}, \ and\ \bibinfo {author} {\bibfnamefont {X.~F.}\
  \bibnamefont {Han}},\ }\href {\doibase 10.1103/PhysRevB.94.174407} {\bibfield
   {journal} {\bibinfo  {journal} {Phys. Rev. B}\ }\textbf {\bibinfo {volume}
  {94}},\ \bibinfo {pages} {174407} (\bibinfo {year} {2016})}\BibitemShut
  {NoStop}%
\bibitem [{\citenamefont {Ando}\ \emph {et~al.}(2008)\citenamefont {Ando},
  \citenamefont {Takahashi}, \citenamefont {Harii}, \citenamefont {Sasage},
  \citenamefont {Ieda}, \citenamefont {Maekawa},\ and\ \citenamefont
  {Saitoh}}]{andoElectricManipulationSpin2008}%
  \BibitemOpen
  \bibfield  {author} {\bibinfo {author} {\bibfnamefont {K.}~\bibnamefont
  {Ando}}, \bibinfo {author} {\bibfnamefont {S.}~\bibnamefont {Takahashi}},
  \bibinfo {author} {\bibfnamefont {K.}~\bibnamefont {Harii}}, \bibinfo
  {author} {\bibfnamefont {K.}~\bibnamefont {Sasage}}, \bibinfo {author}
  {\bibfnamefont {J.}~\bibnamefont {Ieda}}, \bibinfo {author} {\bibfnamefont
  {S.}~\bibnamefont {Maekawa}}, \ and\ \bibinfo {author} {\bibfnamefont
  {E.}~\bibnamefont {Saitoh}},\ }\href {\doibase
  10.1103/PhysRevLett.101.036601} {\bibfield  {journal} {\bibinfo  {journal}
  {Phys. Rev. Lett.}\ }\textbf {\bibinfo {volume} {101}},\ \bibinfo {pages}
  {036601} (\bibinfo {year} {2008})}\BibitemShut {NoStop}%
\bibitem [{\citenamefont {Valenzuela}\ and\ \citenamefont
  {Tinkham}(2006)}]{valenzuelaDirectElectronicMeasurement2006}%
  \BibitemOpen
  \bibfield  {author} {\bibinfo {author} {\bibfnamefont {S.~O.}\ \bibnamefont
  {Valenzuela}}\ and\ \bibinfo {author} {\bibfnamefont {M.}~\bibnamefont
  {Tinkham}},\ }\href {\doibase 10.1038/nature04937} {\bibfield  {journal}
  {\bibinfo  {journal} {Nature}\ }\textbf {\bibinfo {volume} {442}},\ \bibinfo
  {pages} {176} (\bibinfo {year} {2006})}\BibitemShut {NoStop}%
\bibitem [{\citenamefont {Zhao}\ \emph {et~al.}(2006)\citenamefont {Zhao},
  \citenamefont {Loren}, \citenamefont {{van Driel}},\ and\ \citenamefont
  {Smirl}}]{zhaoCoherenceControlHall2006}%
  \BibitemOpen
  \bibfield  {author} {\bibinfo {author} {\bibfnamefont {H.}~\bibnamefont
  {Zhao}}, \bibinfo {author} {\bibfnamefont {E.~J.}\ \bibnamefont {Loren}},
  \bibinfo {author} {\bibfnamefont {H.~M.}\ \bibnamefont {{van Driel}}}, \ and\
  \bibinfo {author} {\bibfnamefont {A.~L.}\ \bibnamefont {Smirl}},\ }\href
  {\doibase 10.1103/PhysRevLett.96.246601} {\bibfield  {journal} {\bibinfo
  {journal} {Phys. Rev. Lett.}\ }\textbf {\bibinfo {volume} {96}},\ \bibinfo
  {pages} {246601} (\bibinfo {year} {2006})}\BibitemShut {NoStop}%
\bibitem [{\citenamefont {Saitoh}\ \emph {et~al.}(2006)\citenamefont {Saitoh},
  \citenamefont {Ueda}, \citenamefont {Miyajima},\ and\ \citenamefont
  {Tatara}}]{saitohConversionSpinCurrent2006}%
  \BibitemOpen
  \bibfield  {author} {\bibinfo {author} {\bibfnamefont {E.}~\bibnamefont
  {Saitoh}}, \bibinfo {author} {\bibfnamefont {M.}~\bibnamefont {Ueda}},
  \bibinfo {author} {\bibfnamefont {H.}~\bibnamefont {Miyajima}}, \ and\
  \bibinfo {author} {\bibfnamefont {G.}~\bibnamefont {Tatara}},\ }\href
  {\doibase 10.1063/1.2199473} {\bibfield  {journal} {\bibinfo  {journal}
  {Appl. Phys. Lett.}\ }\textbf {\bibinfo {volume} {88}},\ \bibinfo {pages}
  {182509} (\bibinfo {year} {2006})}\BibitemShut {NoStop}%
\bibitem [{\citenamefont {Sinova}\ \emph {et~al.}(2004)\citenamefont {Sinova},
  \citenamefont {Culcer}, \citenamefont {Niu}, \citenamefont {Sinitsyn},
  \citenamefont {Jungwirth},\ and\ \citenamefont
  {MacDonald}}]{sinovaUniversalIntrinsicSpin2004}%
  \BibitemOpen
  \bibfield  {author} {\bibinfo {author} {\bibfnamefont {J.}~\bibnamefont
  {Sinova}}, \bibinfo {author} {\bibfnamefont {D.}~\bibnamefont {Culcer}},
  \bibinfo {author} {\bibfnamefont {Q.}~\bibnamefont {Niu}}, \bibinfo {author}
  {\bibfnamefont {N.~A.}\ \bibnamefont {Sinitsyn}}, \bibinfo {author}
  {\bibfnamefont {T.}~\bibnamefont {Jungwirth}}, \ and\ \bibinfo {author}
  {\bibfnamefont {A.~H.}\ \bibnamefont {MacDonald}},\ }\href {\doibase
  10.1103/PhysRevLett.92.126603} {\bibfield  {journal} {\bibinfo  {journal}
  {Phys. Rev. Lett.}\ }\textbf {\bibinfo {volume} {92}},\ \bibinfo {pages}
  {126603} (\bibinfo {year} {2004})}\BibitemShut {NoStop}%
\bibitem [{\citenamefont {Guo}\ \emph {et~al.}(2008)\citenamefont {Guo},
  \citenamefont {Murakami}, \citenamefont {Chen},\ and\ \citenamefont
  {Nagaosa}}]{guoIntrinsicSpinHall2008}%
  \BibitemOpen
  \bibfield  {author} {\bibinfo {author} {\bibfnamefont {G.~Y.}\ \bibnamefont
  {Guo}}, \bibinfo {author} {\bibfnamefont {S.}~\bibnamefont {Murakami}},
  \bibinfo {author} {\bibfnamefont {T.-W.}\ \bibnamefont {Chen}}, \ and\
  \bibinfo {author} {\bibfnamefont {N.}~\bibnamefont {Nagaosa}},\ }\href
  {\doibase 10.1103/PhysRevLett.100.096401} {\bibfield  {journal} {\bibinfo
  {journal} {Phys. Rev. Lett.}\ }\textbf {\bibinfo {volume} {100}},\ \bibinfo
  {pages} {096401} (\bibinfo {year} {2008})}\BibitemShut {NoStop}%
\bibitem [{\citenamefont
  {Murakami}(2003)}]{murakamiDissipationlessQuantumSpin2003}%
  \BibitemOpen
  \bibfield  {author} {\bibinfo {author} {\bibfnamefont {S.}~\bibnamefont
  {Murakami}},\ }\href {\doibase 10.1126/science.1087128} {\bibfield  {journal}
  {\bibinfo  {journal} {Science}\ }\textbf {\bibinfo {volume} {301}},\ \bibinfo
  {pages} {1348} (\bibinfo {year} {2003})}\BibitemShut {NoStop}%
\bibitem [{\citenamefont {Sinova}\ \emph {et~al.}(2015)\citenamefont {Sinova},
  \citenamefont {Valenzuela}, \citenamefont {Wunderlich}, \citenamefont
  {Back},\ and\ \citenamefont {Jungwirth}}]{sinovaSpinHallEffects2015}%
  \BibitemOpen
  \bibfield  {author} {\bibinfo {author} {\bibfnamefont {J.}~\bibnamefont
  {Sinova}}, \bibinfo {author} {\bibfnamefont {S.~O.}\ \bibnamefont
  {Valenzuela}}, \bibinfo {author} {\bibfnamefont {J.}~\bibnamefont
  {Wunderlich}}, \bibinfo {author} {\bibfnamefont {C.~H.}\ \bibnamefont
  {Back}}, \ and\ \bibinfo {author} {\bibfnamefont {T.}~\bibnamefont
  {Jungwirth}},\ }\href {\doibase 10.1103/RevModPhys.87.1213} {\bibfield
  {journal} {\bibinfo  {journal} {Rev. Mod. Phys.}\ }\textbf {\bibinfo {volume}
  {87}},\ \bibinfo {pages} {1213} (\bibinfo {year} {2015})}\BibitemShut
  {NoStop}%
\bibitem [{\citenamefont {Smit}(1955)}]{smitSpontaneousHallEffect1955}%
  \BibitemOpen
  \bibfield  {author} {\bibinfo {author} {\bibfnamefont {J.}~\bibnamefont
  {Smit}},\ }\href {\doibase 10.1016/S0031-8914(55)92596-9} {\bibfield
  {journal} {\bibinfo  {journal} {Physica}\ }\textbf {\bibinfo {volume} {21}},\
  \bibinfo {pages} {877} (\bibinfo {year} {1955})}\BibitemShut {NoStop}%
\bibitem [{\citenamefont {Smit}(1958)}]{smitSpontaneousHallEffect1958}%
  \BibitemOpen
  \bibfield  {author} {\bibinfo {author} {\bibfnamefont {J.}~\bibnamefont
  {Smit}},\ }\href {\doibase 10.1016/S0031-8914(58)93541-9} {\bibfield
  {journal} {\bibinfo  {journal} {Physica}\ }\textbf {\bibinfo {volume} {24}},\
  \bibinfo {pages} {39} (\bibinfo {year} {1958})}\BibitemShut {NoStop}%
\bibitem [{\citenamefont {Berger}(1970)}]{bergerSideJumpMechanismHall1970}%
  \BibitemOpen
  \bibfield  {author} {\bibinfo {author} {\bibfnamefont {L.}~\bibnamefont
  {Berger}},\ }\href {\doibase 10.1103/PhysRevB.2.4559} {\bibfield  {journal}
  {\bibinfo  {journal} {Phys. Rev. B}\ }\textbf {\bibinfo {volume} {2}},\
  \bibinfo {pages} {4559} (\bibinfo {year} {1970})}\BibitemShut {NoStop}%
\bibitem [{\citenamefont {Karplus}\ and\ \citenamefont
  {Luttinger}(1954)}]{karplusHallEffectFerromagnetics1954}%
  \BibitemOpen
  \bibfield  {author} {\bibinfo {author} {\bibfnamefont {R.}~\bibnamefont
  {Karplus}}\ and\ \bibinfo {author} {\bibfnamefont {J.~M.}\ \bibnamefont
  {Luttinger}},\ }\href {\doibase 10.1103/PhysRev.95.1154} {\bibfield
  {journal} {\bibinfo  {journal} {Phys. Rev.}\ }\textbf {\bibinfo {volume}
  {95}},\ \bibinfo {pages} {1154} (\bibinfo {year} {1954})}\BibitemShut
  {NoStop}%
\bibitem [{\citenamefont {Jungwirth}\ \emph {et~al.}(2002)\citenamefont
  {Jungwirth}, \citenamefont {Niu},\ and\ \citenamefont
  {MacDonald}}]{jungwirthAnomalousHallEffect2002}%
  \BibitemOpen
  \bibfield  {author} {\bibinfo {author} {\bibfnamefont {T.}~\bibnamefont
  {Jungwirth}}, \bibinfo {author} {\bibfnamefont {Q.}~\bibnamefont {Niu}}, \
  and\ \bibinfo {author} {\bibfnamefont {A.~H.}\ \bibnamefont {MacDonald}},\
  }\href {\doibase 10.1103/PhysRevLett.88.207208} {\bibfield  {journal}
  {\bibinfo  {journal} {Phys. Rev. Lett.}\ }\textbf {\bibinfo {volume} {88}},\
  \bibinfo {pages} {207208} (\bibinfo {year} {2002})}\BibitemShut {NoStop}%
\bibitem [{\citenamefont {Swihart}\ \emph {et~al.}(1986)\citenamefont
  {Swihart}, \citenamefont {Butler}, \citenamefont {Stocks}, \citenamefont
  {Nicholson},\ and\ \citenamefont
  {Ward}}]{swihartFirstPrinciplesCalculationResidual1986}%
  \BibitemOpen
  \bibfield  {author} {\bibinfo {author} {\bibfnamefont {J.~C.}\ \bibnamefont
  {Swihart}}, \bibinfo {author} {\bibfnamefont {W.~H.}\ \bibnamefont {Butler}},
  \bibinfo {author} {\bibfnamefont {G.~M.}\ \bibnamefont {Stocks}}, \bibinfo
  {author} {\bibfnamefont {D.~M.}\ \bibnamefont {Nicholson}}, \ and\ \bibinfo
  {author} {\bibfnamefont {R.~C.}\ \bibnamefont {Ward}},\ }\href {\doibase
  10.1103/PhysRevLett.57.1181} {\bibfield  {journal} {\bibinfo  {journal}
  {Phys. Rev. Lett.}\ }\textbf {\bibinfo {volume} {57}},\ \bibinfo {pages}
  {1181} (\bibinfo {year} {1986})}\BibitemShut {NoStop}%
\bibitem [{\citenamefont {Zahn}\ \emph {et~al.}(2003)\citenamefont {Zahn},
  \citenamefont {Binder},\ and\ \citenamefont
  {Mertig}}]{zahnImpurityScatteringQuantum2003}%
  \BibitemOpen
  \bibfield  {author} {\bibinfo {author} {\bibfnamefont {P.}~\bibnamefont
  {Zahn}}, \bibinfo {author} {\bibfnamefont {J.}~\bibnamefont {Binder}}, \ and\
  \bibinfo {author} {\bibfnamefont {I.}~\bibnamefont {Mertig}},\ }\href
  {\doibase 10.1103/PhysRevB.68.100403} {\bibfield  {journal} {\bibinfo
  {journal} {Phys. Rev. B}\ }\textbf {\bibinfo {volume} {68}},\ \bibinfo
  {pages} {100403} (\bibinfo {year} {2003})}\BibitemShut {NoStop}%
\bibitem [{\citenamefont {Gradhand}\ \emph
  {et~al.}(2010{\natexlab{a}})\citenamefont {Gradhand}, \citenamefont
  {Fedorov}, \citenamefont {Zahn},\ and\ \citenamefont
  {Mertig}}]{gradhandExtrinsicSpinHall2010}%
  \BibitemOpen
  \bibfield  {author} {\bibinfo {author} {\bibfnamefont {M.}~\bibnamefont
  {Gradhand}}, \bibinfo {author} {\bibfnamefont {D.~V.}\ \bibnamefont
  {Fedorov}}, \bibinfo {author} {\bibfnamefont {P.}~\bibnamefont {Zahn}}, \
  and\ \bibinfo {author} {\bibfnamefont {I.}~\bibnamefont {Mertig}},\ }\href
  {\doibase 10.1103/PhysRevLett.104.186403} {\bibfield  {journal} {\bibinfo
  {journal} {Phys. Rev. Lett.}\ }\textbf {\bibinfo {volume} {104}},\ \bibinfo
  {pages} {186403} (\bibinfo {year} {2010}{\natexlab{a}})}\BibitemShut
  {NoStop}%
\bibitem [{\citenamefont {Guo}\ \emph {et~al.}(2005)\citenamefont {Guo},
  \citenamefont {Yao},\ and\ \citenamefont
  {Niu}}]{guoInitioCalculationIntrinsic2005}%
  \BibitemOpen
  \bibfield  {author} {\bibinfo {author} {\bibfnamefont {G.~Y.}\ \bibnamefont
  {Guo}}, \bibinfo {author} {\bibfnamefont {Y.}~\bibnamefont {Yao}}, \ and\
  \bibinfo {author} {\bibfnamefont {Q.}~\bibnamefont {Niu}},\ }\href {\doibase
  10.1103/PhysRevLett.94.226601} {\bibfield  {journal} {\bibinfo  {journal}
  {Phys. Rev. Lett.}\ }\textbf {\bibinfo {volume} {94}},\ \bibinfo {pages}
  {226601} (\bibinfo {year} {2005})}\BibitemShut {NoStop}%
\bibitem [{\citenamefont {Yao}\ and\ \citenamefont
  {Fang}(2005)}]{yaoSignChangesIntrinsic2005}%
  \BibitemOpen
  \bibfield  {author} {\bibinfo {author} {\bibfnamefont {Y.}~\bibnamefont
  {Yao}}\ and\ \bibinfo {author} {\bibfnamefont {Z.}~\bibnamefont {Fang}},\
  }\href {\doibase 10.1103/PhysRevLett.95.156601} {\bibfield  {journal}
  {\bibinfo  {journal} {Phys. Rev. Lett.}\ }\textbf {\bibinfo {volume} {95}},\
  \bibinfo {pages} {156601} (\bibinfo {year} {2005})}\BibitemShut {NoStop}%
\bibitem [{\citenamefont {Lowitzer}\ \emph {et~al.}(2011)\citenamefont
  {Lowitzer}, \citenamefont {Gradhand}, \citenamefont {K{\"o}dderitzsch},
  \citenamefont {Fedorov}, \citenamefont {Mertig},\ and\ \citenamefont
  {Ebert}}]{lowitzerExtrinsicIntrinsicContributions2011}%
  \BibitemOpen
  \bibfield  {author} {\bibinfo {author} {\bibfnamefont {S.}~\bibnamefont
  {Lowitzer}}, \bibinfo {author} {\bibfnamefont {M.}~\bibnamefont {Gradhand}},
  \bibinfo {author} {\bibfnamefont {D.}~\bibnamefont {K{\"o}dderitzsch}},
  \bibinfo {author} {\bibfnamefont {D.~V.}\ \bibnamefont {Fedorov}}, \bibinfo
  {author} {\bibfnamefont {I.}~\bibnamefont {Mertig}}, \ and\ \bibinfo {author}
  {\bibfnamefont {H.}~\bibnamefont {Ebert}},\ }\href {\doibase
  10.1103/PhysRevLett.106.056601} {\bibfield  {journal} {\bibinfo  {journal}
  {Phys. Rev. Lett.}\ }\textbf {\bibinfo {volume} {106}},\ \bibinfo {pages}
  {056601} (\bibinfo {year} {2011})}\BibitemShut {NoStop}%
\bibitem [{\citenamefont {Guo}(2009)}]{guoInitioCalculationIntrinsic2009}%
  \BibitemOpen
  \bibfield  {author} {\bibinfo {author} {\bibfnamefont {G.~Y.}\ \bibnamefont
  {Guo}},\ }\href@noop {} {\bibfield  {journal} {\bibinfo  {journal} {J. Appl.
  Phys.}\ }\textbf {\bibinfo {volume} {105}},\ \bibinfo {pages} {07C701}
  (\bibinfo {year} {2009})}\BibitemShut {NoStop}%
\bibitem [{\citenamefont {Wang}\ \emph {et~al.}(2006)\citenamefont {Wang},
  \citenamefont {Yates}, \citenamefont {Souza},\ and\ \citenamefont
  {Vanderbilt}}]{wangInitioCalculationAnomalous2006}%
  \BibitemOpen
  \bibfield  {author} {\bibinfo {author} {\bibfnamefont {X.}~\bibnamefont
  {Wang}}, \bibinfo {author} {\bibfnamefont {J.~R.}\ \bibnamefont {Yates}},
  \bibinfo {author} {\bibfnamefont {I.}~\bibnamefont {Souza}}, \ and\ \bibinfo
  {author} {\bibfnamefont {D.}~\bibnamefont {Vanderbilt}},\ }\href {\doibase
  10.1103/PhysRevB.74.195118} {\bibfield  {journal} {\bibinfo  {journal} {Phys.
  Rev. B}\ }\textbf {\bibinfo {volume} {74}},\ \bibinfo {pages} {195118}
  (\bibinfo {year} {2006})}\BibitemShut {NoStop}%
\bibitem [{\citenamefont {Yao}\ \emph {et~al.}(2004)\citenamefont {Yao},
  \citenamefont {Kleinman}, \citenamefont {MacDonald}, \citenamefont {Sinova},
  \citenamefont {Jungwirth}, \citenamefont {Wang}, \citenamefont {Wang},\ and\
  \citenamefont {Niu}}]{yaoFirstPrinciplesCalculation2004}%
  \BibitemOpen
  \bibfield  {author} {\bibinfo {author} {\bibfnamefont {Y.}~\bibnamefont
  {Yao}}, \bibinfo {author} {\bibfnamefont {L.}~\bibnamefont {Kleinman}},
  \bibinfo {author} {\bibfnamefont {A.~H.}\ \bibnamefont {MacDonald}}, \bibinfo
  {author} {\bibfnamefont {J.}~\bibnamefont {Sinova}}, \bibinfo {author}
  {\bibfnamefont {T.}~\bibnamefont {Jungwirth}}, \bibinfo {author}
  {\bibfnamefont {D.-s.}\ \bibnamefont {Wang}}, \bibinfo {author}
  {\bibfnamefont {E.}~\bibnamefont {Wang}}, \ and\ \bibinfo {author}
  {\bibfnamefont {Q.}~\bibnamefont {Niu}},\ }\href {\doibase
  10.1103/PhysRevLett.92.037204} {\bibfield  {journal} {\bibinfo  {journal}
  {Phys. Rev. Lett.}\ }\textbf {\bibinfo {volume} {92}},\ \bibinfo {pages}
  {037204} (\bibinfo {year} {2004})}\BibitemShut {NoStop}%
\bibitem [{\citenamefont {Stamm}\ \emph {et~al.}(2017)\citenamefont {Stamm},
  \citenamefont {Murer}, \citenamefont {Berritta}, \citenamefont {Feng},
  \citenamefont {Gabureac}, \citenamefont {Oppeneer},\ and\ \citenamefont
  {Gambardella}}]{stammMagnetoOpticalDetectionSpin2017}%
  \BibitemOpen
  \bibfield  {author} {\bibinfo {author} {\bibfnamefont {C.}~\bibnamefont
  {Stamm}}, \bibinfo {author} {\bibfnamefont {C.}~\bibnamefont {Murer}},
  \bibinfo {author} {\bibfnamefont {M.}~\bibnamefont {Berritta}}, \bibinfo
  {author} {\bibfnamefont {J.}~\bibnamefont {Feng}}, \bibinfo {author}
  {\bibfnamefont {M.}~\bibnamefont {Gabureac}}, \bibinfo {author}
  {\bibfnamefont {P.~M.}\ \bibnamefont {Oppeneer}}, \ and\ \bibinfo {author}
  {\bibfnamefont {P.}~\bibnamefont {Gambardella}},\ }\href {\doibase
  10.1103/PhysRevLett.119.087203} {\bibfield  {journal} {\bibinfo  {journal}
  {Phys. Rev. Lett.}\ }\textbf {\bibinfo {volume} {119}},\ \bibinfo {pages}
  {087203} (\bibinfo {year} {2017})}\BibitemShut {NoStop}%
\bibitem [{\citenamefont {Zhang}(2000)}]{zhangSpinHallEffect2000}%
  \BibitemOpen
  \bibfield  {author} {\bibinfo {author} {\bibfnamefont {S.}~\bibnamefont
  {Zhang}},\ }\href {\doibase 10.1103/PhysRevLett.85.393} {\bibfield  {journal}
  {\bibinfo  {journal} {Phys. Rev. Lett.}\ }\textbf {\bibinfo {volume} {85}},\
  \bibinfo {pages} {393} (\bibinfo {year} {2000})}\BibitemShut {NoStop}%
\bibitem [{\citenamefont {Freimuth}\ \emph {et~al.}(2014)\citenamefont
  {Freimuth}, \citenamefont {Bl{\"u}gel},\ and\ \citenamefont
  {Mokrousov}}]{freimuthSpinorbitTorquesCo2014}%
  \BibitemOpen
  \bibfield  {author} {\bibinfo {author} {\bibfnamefont {F.}~\bibnamefont
  {Freimuth}}, \bibinfo {author} {\bibfnamefont {S.}~\bibnamefont
  {Bl{\"u}gel}}, \ and\ \bibinfo {author} {\bibfnamefont {Y.}~\bibnamefont
  {Mokrousov}},\ }\href {\doibase 10.1103/PhysRevB.90.174423} {\bibfield
  {journal} {\bibinfo  {journal} {Phys. Rev. B}\ }\textbf {\bibinfo {volume}
  {90}},\ \bibinfo {pages} {174423} (\bibinfo {year} {2014})}\BibitemShut
  {NoStop}%
\bibitem [{\citenamefont {Wimmer}\ \emph {et~al.}(2016)\citenamefont {Wimmer},
  \citenamefont {Chadova}, \citenamefont {Seemann}, \citenamefont
  {K{\"o}dderitzsch},\ and\ \citenamefont
  {Ebert}}]{wimmerFullyRelativisticDescription2016}%
  \BibitemOpen
  \bibfield  {author} {\bibinfo {author} {\bibfnamefont {S.}~\bibnamefont
  {Wimmer}}, \bibinfo {author} {\bibfnamefont {K.}~\bibnamefont {Chadova}},
  \bibinfo {author} {\bibfnamefont {M.}~\bibnamefont {Seemann}}, \bibinfo
  {author} {\bibfnamefont {D.}~\bibnamefont {K{\"o}dderitzsch}}, \ and\
  \bibinfo {author} {\bibfnamefont {H.}~\bibnamefont {Ebert}},\ }\href
  {\doibase 10.1103/PhysRevB.94.054415} {\bibfield  {journal} {\bibinfo
  {journal} {Phys. Rev. B}\ }\textbf {\bibinfo {volume} {94}},\ \bibinfo
  {pages} {054415} (\bibinfo {year} {2016})}\BibitemShut {NoStop}%
\bibitem [{\citenamefont {G{\'e}ranton}\ \emph {et~al.}(2016)\citenamefont
  {G{\'e}ranton}, \citenamefont {Zimmermann}, \citenamefont {Long},
  \citenamefont {Mavropoulos}, \citenamefont {Bl{\"u}gel}, \citenamefont
  {Freimuth},\ and\ \citenamefont
  {Mokrousov}}]{gerantonSpinorbitTorquesSpin2016}%
  \BibitemOpen
  \bibfield  {author} {\bibinfo {author} {\bibfnamefont {G.}~\bibnamefont
  {G{\'e}ranton}}, \bibinfo {author} {\bibfnamefont {B.}~\bibnamefont
  {Zimmermann}}, \bibinfo {author} {\bibfnamefont {N.~H.}\ \bibnamefont
  {Long}}, \bibinfo {author} {\bibfnamefont {P.}~\bibnamefont {Mavropoulos}},
  \bibinfo {author} {\bibfnamefont {S.}~\bibnamefont {Bl{\"u}gel}}, \bibinfo
  {author} {\bibfnamefont {F.}~\bibnamefont {Freimuth}}, \ and\ \bibinfo
  {author} {\bibfnamefont {Y.}~\bibnamefont {Mokrousov}},\ }\href {\doibase
  10.1103/PhysRevB.93.224420} {\bibfield  {journal} {\bibinfo  {journal} {Phys.
  Rev. B}\ }\textbf {\bibinfo {volume} {93}},\ \bibinfo {pages} {224420}
  (\bibinfo {year} {2016})}\BibitemShut {NoStop}%
\bibitem [{\citenamefont {K{\"o}dderitzsch}\ \emph {et~al.}(2015)\citenamefont
  {K{\"o}dderitzsch}, \citenamefont {Chadova},\ and\ \citenamefont
  {Ebert}}]{kodderitzschLinearResponseKuboBastin2015}%
  \BibitemOpen
  \bibfield  {author} {\bibinfo {author} {\bibfnamefont {D.}~\bibnamefont
  {K{\"o}dderitzsch}}, \bibinfo {author} {\bibfnamefont {K.}~\bibnamefont
  {Chadova}}, \ and\ \bibinfo {author} {\bibfnamefont {H.}~\bibnamefont
  {Ebert}},\ }\href {\doibase 10.1103/PhysRevB.92.184415} {\bibfield  {journal}
  {\bibinfo  {journal} {Phys. Rev. B}\ }\textbf {\bibinfo {volume} {92}},\
  \bibinfo {pages} {184415} (\bibinfo {year} {2015})}\BibitemShut {NoStop}%
\bibitem [{\citenamefont {Kosma}\ \emph {et~al.}(2020)\citenamefont {Kosma},
  \citenamefont {R{\"u}{\ss}mann}, \citenamefont {Bl{\"u}gel},\ and\
  \citenamefont {Mavropoulos}}]{kosmaStrongSpinorbitTorque2020}%
  \BibitemOpen
  \bibfield  {author} {\bibinfo {author} {\bibfnamefont {A.}~\bibnamefont
  {Kosma}}, \bibinfo {author} {\bibfnamefont {P.}~\bibnamefont
  {R{\"u}{\ss}mann}}, \bibinfo {author} {\bibfnamefont {S.}~\bibnamefont
  {Bl{\"u}gel}}, \ and\ \bibinfo {author} {\bibfnamefont {P.}~\bibnamefont
  {Mavropoulos}},\ }\href {\doibase 10.1103/PhysRevB.102.144424} {\bibfield
  {journal} {\bibinfo  {journal} {Phys. Rev. B}\ }\textbf {\bibinfo {volume}
  {102}},\ \bibinfo {pages} {144424} (\bibinfo {year} {2020})}\BibitemShut
  {NoStop}%
\bibitem [{\citenamefont {Yang}\ \emph {et~al.}(2016)\citenamefont {Yang},
  \citenamefont {Cai}, \citenamefont {Ju}, \citenamefont {Edmonds},
  \citenamefont {Yang}, \citenamefont {Liu}, \citenamefont {Li}, \citenamefont
  {Zhang}, \citenamefont {Sheng}, \citenamefont {Wang}, \citenamefont {Ji},\
  and\ \citenamefont {Wang}}]{yangSpinorbitTorquePt2016}%
  \BibitemOpen
  \bibfield  {author} {\bibinfo {author} {\bibfnamefont {M.}~\bibnamefont
  {Yang}}, \bibinfo {author} {\bibfnamefont {K.}~\bibnamefont {Cai}}, \bibinfo
  {author} {\bibfnamefont {H.}~\bibnamefont {Ju}}, \bibinfo {author}
  {\bibfnamefont {K.~W.}\ \bibnamefont {Edmonds}}, \bibinfo {author}
  {\bibfnamefont {G.}~\bibnamefont {Yang}}, \bibinfo {author} {\bibfnamefont
  {S.}~\bibnamefont {Liu}}, \bibinfo {author} {\bibfnamefont {B.}~\bibnamefont
  {Li}}, \bibinfo {author} {\bibfnamefont {B.}~\bibnamefont {Zhang}}, \bibinfo
  {author} {\bibfnamefont {Y.}~\bibnamefont {Sheng}}, \bibinfo {author}
  {\bibfnamefont {S.}~\bibnamefont {Wang}}, \bibinfo {author} {\bibfnamefont
  {Y.}~\bibnamefont {Ji}}, \ and\ \bibinfo {author} {\bibfnamefont
  {K.}~\bibnamefont {Wang}},\ }\href {\doibase 10.1038/srep20778} {\bibfield
  {journal} {\bibinfo  {journal} {Sci. Rep.}\ }\textbf {\bibinfo {volume}
  {6}},\ \bibinfo {pages} {20778} (\bibinfo {year} {2016})}\BibitemShut
  {NoStop}%
\bibitem [{\citenamefont {Avci}\ \emph {et~al.}(2014)\citenamefont {Avci},
  \citenamefont {Garello}, \citenamefont {Nistor}, \citenamefont {Godey},
  \citenamefont {Ballesteros}, \citenamefont {Mugarza}, \citenamefont {Barla},
  \citenamefont {Valvidares}, \citenamefont {Pellegrin}, \citenamefont {Ghosh},
  \citenamefont {Miron}, \citenamefont {Boulle}, \citenamefont {Auffret},
  \citenamefont {Gaudin},\ and\ \citenamefont
  {Gambardella}}]{avciFieldlikeAntidampingSpinorbit2014}%
  \BibitemOpen
  \bibfield  {author} {\bibinfo {author} {\bibfnamefont {C.~O.}\ \bibnamefont
  {Avci}}, \bibinfo {author} {\bibfnamefont {K.}~\bibnamefont {Garello}},
  \bibinfo {author} {\bibfnamefont {C.}~\bibnamefont {Nistor}}, \bibinfo
  {author} {\bibfnamefont {S.}~\bibnamefont {Godey}}, \bibinfo {author}
  {\bibfnamefont {B.}~\bibnamefont {Ballesteros}}, \bibinfo {author}
  {\bibfnamefont {A.}~\bibnamefont {Mugarza}}, \bibinfo {author} {\bibfnamefont
  {A.}~\bibnamefont {Barla}}, \bibinfo {author} {\bibfnamefont
  {M.}~\bibnamefont {Valvidares}}, \bibinfo {author} {\bibfnamefont
  {E.}~\bibnamefont {Pellegrin}}, \bibinfo {author} {\bibfnamefont
  {A.}~\bibnamefont {Ghosh}}, \bibinfo {author} {\bibfnamefont {I.~M.}\
  \bibnamefont {Miron}}, \bibinfo {author} {\bibfnamefont {O.}~\bibnamefont
  {Boulle}}, \bibinfo {author} {\bibfnamefont {S.}~\bibnamefont {Auffret}},
  \bibinfo {author} {\bibfnamefont {G.}~\bibnamefont {Gaudin}}, \ and\ \bibinfo
  {author} {\bibfnamefont {P.}~\bibnamefont {Gambardella}},\ }\href {\doibase
  10.1103/PhysRevB.89.214419} {\bibfield  {journal} {\bibinfo  {journal} {Phys.
  Rev. B}\ }\textbf {\bibinfo {volume} {89}},\ \bibinfo {pages} {214419}
  (\bibinfo {year} {2014})}\BibitemShut {NoStop}%
\bibitem [{\citenamefont {Nikoli{\'c}}\ \emph {et~al.}(2005)\citenamefont
  {Nikoli{\'c}}, \citenamefont {Souma}, \citenamefont {Z{\^a}rbo},\ and\
  \citenamefont {Sinova}}]{nikolicNonequilibriumSpinHall2005}%
  \BibitemOpen
  \bibfield  {author} {\bibinfo {author} {\bibfnamefont {B.~K.}\ \bibnamefont
  {Nikoli{\'c}}}, \bibinfo {author} {\bibfnamefont {S.}~\bibnamefont {Souma}},
  \bibinfo {author} {\bibfnamefont {L.~P.}\ \bibnamefont {Z{\^a}rbo}}, \ and\
  \bibinfo {author} {\bibfnamefont {J.}~\bibnamefont {Sinova}},\ }\href
  {\doibase 10.1103/PhysRevLett.95.046601} {\bibfield  {journal} {\bibinfo
  {journal} {Phys. Rev. Lett.}\ }\textbf {\bibinfo {volume} {95}},\ \bibinfo
  {pages} {046601} (\bibinfo {year} {2005})},\ \Eprint
  {http://arxiv.org/abs/cond-mat/0412595} {arXiv:cond-mat/0412595} \BibitemShut
  {NoStop}%
\bibitem [{\citenamefont {Nikolic}\ \emph {et~al.}(2006)\citenamefont
  {Nikolic}, \citenamefont {Zarbo},\ and\ \citenamefont
  {Souma}}]{nikolicImagingMesoscopicSpin2006}%
  \BibitemOpen
  \bibfield  {author} {\bibinfo {author} {\bibfnamefont {B.~K.}\ \bibnamefont
  {Nikolic}}, \bibinfo {author} {\bibfnamefont {L.~P.}\ \bibnamefont {Zarbo}},
  \ and\ \bibinfo {author} {\bibfnamefont {S.}~\bibnamefont {Souma}},\ }\href
  {\doibase 10.1103/PhysRevB.73.075303} {\bibfield  {journal} {\bibinfo
  {journal} {Phys. Rev. B}\ }\textbf {\bibinfo {volume} {73}},\ \bibinfo
  {pages} {075303} (\bibinfo {year} {2006})},\ \Eprint
  {http://arxiv.org/abs/cond-mat/0506588} {arXiv:cond-mat/0506588} \BibitemShut
  {NoStop}%
\bibitem [{\citenamefont
  {Zabloudil}(2005)}]{zabloudilElectronScatteringSolid2005}%
  \BibitemOpen
  \bibinfo {editor} {\bibfnamefont {J.}~\bibnamefont {Zabloudil}},\ ed.,\
  \href@noop {} {\emph {\bibinfo {title} {Electron Scattering in Solid Matter:
  A Theoretical and Computational Treatise}}},\ \bibinfo {series} {Springer
  Series in Solid-State Sciences}\ No.\ \bibinfo {number} {147}\ (\bibinfo
  {publisher} {{Springer}},\ \bibinfo {address} {{Berlin ; New York}},\
  \bibinfo {year} {2005})\BibitemShut {NoStop}%
\bibitem [{\citenamefont {Gradhand}\ \emph {et~al.}(2009)\citenamefont
  {Gradhand}, \citenamefont {Czerner}, \citenamefont {Fedorov}, \citenamefont
  {Zahn}, \citenamefont {Yavorsky}, \citenamefont {Szunyogh},\ and\
  \citenamefont {Mertig}}]{gradhandSpinPolarizationFermi2009}%
  \BibitemOpen
  \bibfield  {author} {\bibinfo {author} {\bibfnamefont {M.}~\bibnamefont
  {Gradhand}}, \bibinfo {author} {\bibfnamefont {M.}~\bibnamefont {Czerner}},
  \bibinfo {author} {\bibfnamefont {D.~V.}\ \bibnamefont {Fedorov}}, \bibinfo
  {author} {\bibfnamefont {P.}~\bibnamefont {Zahn}}, \bibinfo {author}
  {\bibfnamefont {B.~Y.}\ \bibnamefont {Yavorsky}}, \bibinfo {author}
  {\bibfnamefont {L.}~\bibnamefont {Szunyogh}}, \ and\ \bibinfo {author}
  {\bibfnamefont {I.}~\bibnamefont {Mertig}},\ }\href {\doibase
  10.1103/PhysRevB.80.224413} {\bibfield  {journal} {\bibinfo  {journal} {Phys.
  Rev. B}\ }\textbf {\bibinfo {volume} {80}},\ \bibinfo {pages} {224413}
  (\bibinfo {year} {2009})}\BibitemShut {NoStop}%
\bibitem [{\citenamefont {Gradhand}\ \emph
  {et~al.}(2010{\natexlab{b}})\citenamefont {Gradhand}, \citenamefont
  {Fedorov}, \citenamefont {Zahn},\ and\ \citenamefont
  {Mertig}}]{gradhandFullyRelativisticInitio2010}%
  \BibitemOpen
  \bibfield  {author} {\bibinfo {author} {\bibfnamefont {M.}~\bibnamefont
  {Gradhand}}, \bibinfo {author} {\bibfnamefont {D.~V.}\ \bibnamefont
  {Fedorov}}, \bibinfo {author} {\bibfnamefont {P.}~\bibnamefont {Zahn}}, \
  and\ \bibinfo {author} {\bibfnamefont {I.}~\bibnamefont {Mertig}},\ }\href
  {\doibase 10.1103/PhysRevB.81.020403} {\bibfield  {journal} {\bibinfo
  {journal} {Phys. Rev. B}\ }\textbf {\bibinfo {volume} {81}},\ \bibinfo
  {pages} {020403} (\bibinfo {year} {2010}{\natexlab{b}})}\BibitemShut
  {NoStop}%
\bibitem [{\citenamefont {Heiliger}\ \emph {et~al.}(2008)\citenamefont
  {Heiliger}, \citenamefont {Czerner}, \citenamefont {Yavorsky}, \citenamefont
  {Mertig},\ and\ \citenamefont
  {Stiles}}]{heiligerImplementationNonequilibriumGreen2008}%
  \BibitemOpen
  \bibfield  {author} {\bibinfo {author} {\bibfnamefont {C.}~\bibnamefont
  {Heiliger}}, \bibinfo {author} {\bibfnamefont {M.}~\bibnamefont {Czerner}},
  \bibinfo {author} {\bibfnamefont {B.~Y.}\ \bibnamefont {Yavorsky}}, \bibinfo
  {author} {\bibfnamefont {I.}~\bibnamefont {Mertig}}, \ and\ \bibinfo {author}
  {\bibfnamefont {M.~D.}\ \bibnamefont {Stiles}},\ }\href {\doibase
  10.1063/1.2835071} {\bibfield  {journal} {\bibinfo  {journal} {J. Appl.
  Phys.}\ }\textbf {\bibinfo {volume} {103}},\ \bibinfo {pages} {07A709}
  (\bibinfo {year} {2008})}\BibitemShut {NoStop}%
\bibitem [{\citenamefont {Franz}\ \emph {et~al.}(2013)\citenamefont {Franz},
  \citenamefont {Czerner},\ and\ \citenamefont
  {Heiliger}}]{franzImplementationNonequilibriumVertex2013}%
  \BibitemOpen
  \bibfield  {author} {\bibinfo {author} {\bibfnamefont {C.}~\bibnamefont
  {Franz}}, \bibinfo {author} {\bibfnamefont {M.}~\bibnamefont {Czerner}}, \
  and\ \bibinfo {author} {\bibfnamefont {C.}~\bibnamefont {Heiliger}},\ }\href
  {\doibase 10.1088/0953-8984/25/42/425301} {\bibfield  {journal} {\bibinfo
  {journal} {J. Phys. Condens. Matter}\ }\textbf {\bibinfo {volume} {25}},\
  \bibinfo {pages} {425301} (\bibinfo {year} {2013})}\BibitemShut {NoStop}%
\bibitem [{\citenamefont {Mahr}\ \emph {et~al.}(2017)\citenamefont {Mahr},
  \citenamefont {Czerner},\ and\ \citenamefont
  {Heiliger}}]{mahrImplementationMethodCalculating2017}%
  \BibitemOpen
  \bibfield  {author} {\bibinfo {author} {\bibfnamefont {C.~E.}\ \bibnamefont
  {Mahr}}, \bibinfo {author} {\bibfnamefont {M.}~\bibnamefont {Czerner}}, \
  and\ \bibinfo {author} {\bibfnamefont {C.}~\bibnamefont {Heiliger}},\ }\href
  {\doibase 10.1103/PhysRevB.96.165121} {\bibfield  {journal} {\bibinfo
  {journal} {Phys. Rev. B}\ }\textbf {\bibinfo {volume} {96}},\ \bibinfo
  {pages} {165121} (\bibinfo {year} {2017})}\BibitemShut {NoStop}%
\bibitem [{\citenamefont
  {Datta}(1995)}]{dattaElectronicTransportMesoscopic1995}%
  \BibitemOpen
  \bibfield  {author} {\bibinfo {author} {\bibfnamefont {S.}~\bibnamefont
  {Datta}},\ }\href {\doibase 10.1017/CBO9780511805776} {\emph {\bibinfo
  {title} {Electronic {{Transport}} in {{Mesoscopic Systems}}}}}\ (\bibinfo
  {publisher} {{Cambridge University Press}},\ \bibinfo {year}
  {1995})\BibitemShut {NoStop}%
\bibitem [{\citenamefont {Herschbach}\ \emph {et~al.}(2012)\citenamefont
  {Herschbach}, \citenamefont {Gradhand}, \citenamefont {Fedorov},\ and\
  \citenamefont {Mertig}}]{herschbachEnhancementSpinHall2012}%
  \BibitemOpen
  \bibfield  {author} {\bibinfo {author} {\bibfnamefont {C.}~\bibnamefont
  {Herschbach}}, \bibinfo {author} {\bibfnamefont {M.}~\bibnamefont
  {Gradhand}}, \bibinfo {author} {\bibfnamefont {D.~V.}\ \bibnamefont
  {Fedorov}}, \ and\ \bibinfo {author} {\bibfnamefont {I.}~\bibnamefont
  {Mertig}},\ }\href {\doibase 10.1103/PhysRevB.85.195133} {\bibfield
  {journal} {\bibinfo  {journal} {Phys. Rev. B}\ }\textbf {\bibinfo {volume}
  {85}},\ \bibinfo {pages} {195133} (\bibinfo {year} {2012})}\BibitemShut
  {NoStop}%
\bibitem [{\citenamefont {Sharvin}(1965)}]{sharvinPossibleMethodStudying1965}%
  \BibitemOpen
  \bibfield  {author} {\bibinfo {author} {\bibfnamefont {Y.~V.}\ \bibnamefont
  {Sharvin}},\ }\href@noop {} {\bibfield  {journal} {\bibinfo  {journal} {Zh.
  Eksperim. i Teor. Fiz.}\ }\textbf {\bibinfo {volume} {48}},\ \bibinfo {pages}
  {984} (\bibinfo {year} {1965})}\BibitemShut {NoStop}%
\bibitem [{\citenamefont {Wang}\ \emph {et~al.}(2014)\citenamefont {Wang},
  \citenamefont {Du}, \citenamefont {Pu}, \citenamefont {Adur}, \citenamefont
  {Hammel},\ and\ \citenamefont {Yang}}]{wangScalingSpinHall2014}%
  \BibitemOpen
  \bibfield  {author} {\bibinfo {author} {\bibfnamefont {H.~L.}\ \bibnamefont
  {Wang}}, \bibinfo {author} {\bibfnamefont {C.~H.}\ \bibnamefont {Du}},
  \bibinfo {author} {\bibfnamefont {Y.}~\bibnamefont {Pu}}, \bibinfo {author}
  {\bibfnamefont {R.}~\bibnamefont {Adur}}, \bibinfo {author} {\bibfnamefont
  {P.~C.}\ \bibnamefont {Hammel}}, \ and\ \bibinfo {author} {\bibfnamefont
  {F.~Y.}\ \bibnamefont {Yang}},\ }\href {\doibase
  10.1103/PhysRevLett.112.197201} {\bibfield  {journal} {\bibinfo  {journal}
  {Phys. Rev. Lett.}\ }\textbf {\bibinfo {volume} {112}},\ \bibinfo {pages}
  {197201} (\bibinfo {year} {2014})}\BibitemShut {NoStop}%
\bibitem [{\citenamefont {Qiao}\ \emph {et~al.}(2018)\citenamefont {Qiao},
  \citenamefont {Zhou}, \citenamefont {Yuan},\ and\ \citenamefont
  {Zhao}}]{qiaoCalculationIntrinsicSpin2018}%
  \BibitemOpen
  \bibfield  {author} {\bibinfo {author} {\bibfnamefont {J.}~\bibnamefont
  {Qiao}}, \bibinfo {author} {\bibfnamefont {J.}~\bibnamefont {Zhou}}, \bibinfo
  {author} {\bibfnamefont {Z.}~\bibnamefont {Yuan}}, \ and\ \bibinfo {author}
  {\bibfnamefont {W.}~\bibnamefont {Zhao}},\ }\href {\doibase
  10.1103/PhysRevB.98.214402} {\bibfield  {journal} {\bibinfo  {journal} {Phys.
  Rev. B}\ }\textbf {\bibinfo {volume} {98}},\ \bibinfo {pages} {214402}
  (\bibinfo {year} {2018})}\BibitemShut {NoStop}%
\bibitem [{\citenamefont {Wu}\ \emph {et~al.}(2020)\citenamefont {Wu},
  \citenamefont {Rossignol},\ and\ \citenamefont
  {Gradhand}}]{wuSpindependentTransportUranium2020}%
  \BibitemOpen
  \bibfield  {author} {\bibinfo {author} {\bibfnamefont {M.-H.}\ \bibnamefont
  {Wu}}, \bibinfo {author} {\bibfnamefont {H.}~\bibnamefont {Rossignol}}, \
  and\ \bibinfo {author} {\bibfnamefont {M.}~\bibnamefont {Gradhand}},\ }\href
  {\doibase 10.1103/PhysRevB.101.224411} {\bibfield  {journal} {\bibinfo
  {journal} {Phys. Rev. B}\ }\textbf {\bibinfo {volume} {101}},\ \bibinfo
  {pages} {224411} (\bibinfo {year} {2020})}\BibitemShut {NoStop}%
\bibitem [{Note1()}]{Note1}%
  \BibitemOpen
  \bibinfo {note} {Supplemental Material}\BibitemShut {NoStop}%
\bibitem [{\citenamefont {Sagasta}\ \emph {et~al.}(2016)\citenamefont
  {Sagasta}, \citenamefont {Omori}, \citenamefont {Isasa}, \citenamefont
  {Gradhand}, \citenamefont {Hueso}, \citenamefont {Niimi}, \citenamefont
  {Otani},\ and\ \citenamefont {Casanova}}]{sagastaTuningSpinHall2016}%
  \BibitemOpen
  \bibfield  {author} {\bibinfo {author} {\bibfnamefont {E.}~\bibnamefont
  {Sagasta}}, \bibinfo {author} {\bibfnamefont {Y.}~\bibnamefont {Omori}},
  \bibinfo {author} {\bibfnamefont {M.}~\bibnamefont {Isasa}}, \bibinfo
  {author} {\bibfnamefont {M.}~\bibnamefont {Gradhand}}, \bibinfo {author}
  {\bibfnamefont {L.~E.}\ \bibnamefont {Hueso}}, \bibinfo {author}
  {\bibfnamefont {Y.}~\bibnamefont {Niimi}}, \bibinfo {author} {\bibfnamefont
  {Y.}~\bibnamefont {Otani}}, \ and\ \bibinfo {author} {\bibfnamefont
  {F.}~\bibnamefont {Casanova}},\ }\href {\doibase 10.1103/PhysRevB.94.060412}
  {\bibfield  {journal} {\bibinfo  {journal} {Phys. Rev. B}\ }\textbf {\bibinfo
  {volume} {94}},\ \bibinfo {pages} {060412} (\bibinfo {year}
  {2016})}\BibitemShut {NoStop}%
\bibitem [{\citenamefont {Skinner}\ \emph {et~al.}(2015)\citenamefont
  {Skinner}, \citenamefont {Olejn{\'i}k}, \citenamefont {Cunningham},
  \citenamefont {Kurebayashi}, \citenamefont {Campion}, \citenamefont
  {Gallagher}, \citenamefont {Jungwirth},\ and\ \citenamefont
  {Ferguson}}]{skinnerComplementarySpinHallInverse2015}%
  \BibitemOpen
  \bibfield  {author} {\bibinfo {author} {\bibfnamefont {T.~D.}\ \bibnamefont
  {Skinner}}, \bibinfo {author} {\bibfnamefont {K.}~\bibnamefont
  {Olejn{\'i}k}}, \bibinfo {author} {\bibfnamefont {L.~K.}\ \bibnamefont
  {Cunningham}}, \bibinfo {author} {\bibfnamefont {H.}~\bibnamefont
  {Kurebayashi}}, \bibinfo {author} {\bibfnamefont {R.~P.}\ \bibnamefont
  {Campion}}, \bibinfo {author} {\bibfnamefont {B.~L.}\ \bibnamefont
  {Gallagher}}, \bibinfo {author} {\bibfnamefont {T.}~\bibnamefont
  {Jungwirth}}, \ and\ \bibinfo {author} {\bibfnamefont {A.~J.}\ \bibnamefont
  {Ferguson}},\ }\href {\doibase 10.1038/ncomms7730} {\bibfield  {journal}
  {\bibinfo  {journal} {Nat. Commun.}\ }\textbf {\bibinfo {volume} {6}},\
  \bibinfo {pages} {6730} (\bibinfo {year} {2015})}\BibitemShut {NoStop}%
\end{thebibliography}%

\end{document}


\preprint{APS/123-QED}

\title{Supplemental Material:\\Spin accumulation from non-equilibrium first principles methods}

\author{Alexander Fabian}
\author{Michael Czerner}
\author{Christian Heiliger}
\affiliation{Institut für Theoretische Physik, Justus-Liebig-Universität Giessen, Heinrich-Buff-Ring 16, 35392 Giessen, Germany}
\affiliation{Center for Materials Research (LaMa), Justus-Liebig-Universität Giessen, Heinrich-Buff-Ring 16, 35392 Giessen, Germany }

\author{Hugo Rossignol}
\affiliation{HH Wills Physics Laboratory, University of Bristol, Tyndall Avenue BS8 1TL, United Kingdom}
\author{Ming-Hung Wu}
\affiliation{HH Wills Physics Laboratory, University of Bristol, Tyndall Avenue BS8 1TL, United Kingdom}
\author{Martin Gradhand}
\affiliation{HH Wills Physics Laboratory, University of Bristol, Tyndall Avenue BS8 1TL, United Kingdom}
\affiliation{Institute of Physics, Johannes Gutenberg University Mainz, 55099 Mainz, Germany}


\begin{abstract}

\end{abstract}

\maketitle

\tableofcontents

\section*{Organization}
The supplemental information is organized as follows: In section I  we show the calculated accumulation for all small systems considered in the main text.
In section II we show the accumulation for larger systems of Cu, Pt, and U. We make a short remark on the behavior of U.
In section III we conclude the supplemental information by showing the Fermi surfaces of the considered systems with colour coded group velocity.
\vfill
\clearpage
\section{Accumulations for small systems}

\begin{figure*}[htbp]
	\includegraphics[width=.47\textwidth]{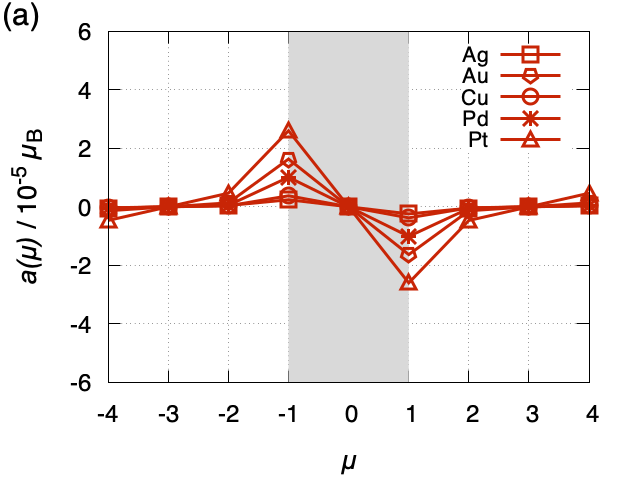}
	\includegraphics[width=.47\textwidth]{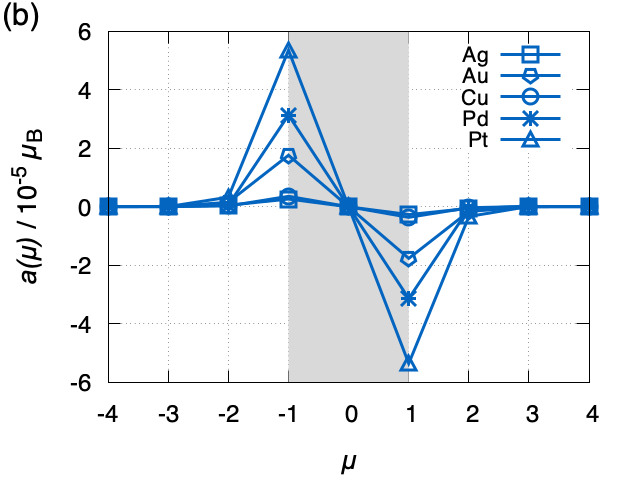}
	\caption{\label{fig:fcc} Plot of the magnetic moment per atom for the fcc systems. (a) Keldysh, (b) Boltzmann. The thin film is highlighted in grey. Each line shows the same antisymmetric behaviour. The agreement of (a) and (b) between Cu, Ag, and Au is better than for Pd and Pt.}
\end{figure*}

\begin{figure*}[htbp]
\includegraphics[width=.47\textwidth]{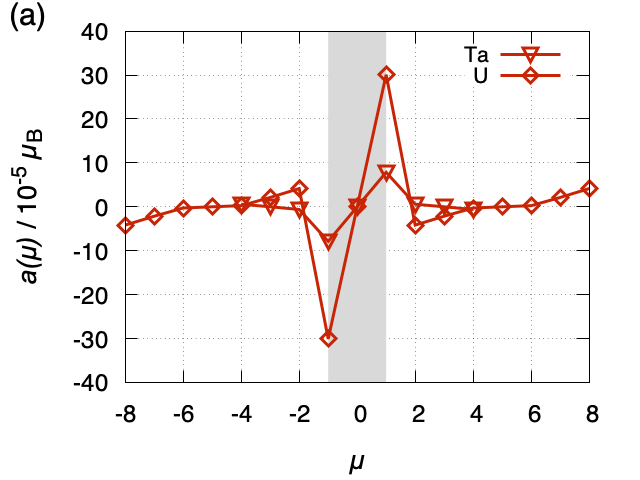}
\includegraphics[width=.47\textwidth]{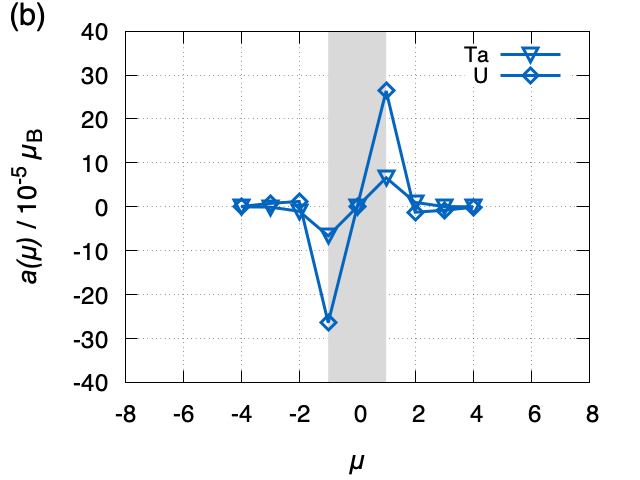}
	\caption{\label{fig:bcc} Plot of the magnetic moment per atom for the bcc systems. (a) Keldysh, (b) Boltzmann. The thin film is highlighted in grey. Each line shows the same antisymmetric behaviour. The sign of the first extremum is changed compared to the systems with fcc lattice. The agreement of the absolute values is not of the same order as for Cu, Ag, and Au.}
\end{figure*}

The accumulation of all the small (3 atomic layers) systems is depicted in Fig.~\ref{fig:fcc} for the fcc (Ag, Au, Cu, Pd, Pt) and Fig.~\ref{fig:bcc} for the bcc systems (Ta, U) for all elements considered in the main text. The blue lines are calculated via the Boltzmann formalism, the red lines are calculated via the Keldysh formalism. The accumulation shows qualitatively the same symmetry enforced anti-symmetric behaviour in both methods. The trend between the elements is quite consistent for both methods. The sign change in the accumulation for U and Ta is consistent in both methods. For a detailed discussion refer to the main text. For an overview of the trend refer to Table~I of the main text.

\clearpage

\section{Accumulations for large systems}
\begin{figure}[htbp]
	\includegraphics[width=.47\textwidth]{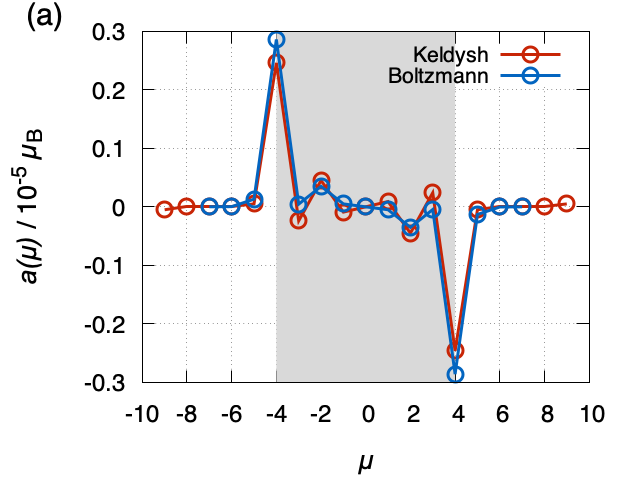}
	\includegraphics[width=.47\textwidth]{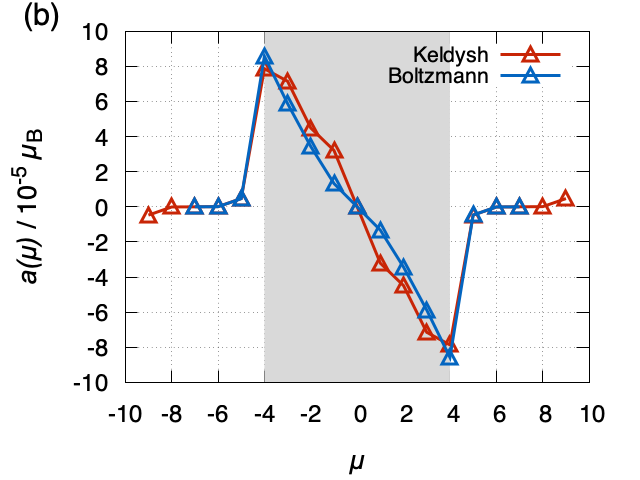}\\
	\vspace{.7cm}
	\includegraphics[width=.47\textwidth]{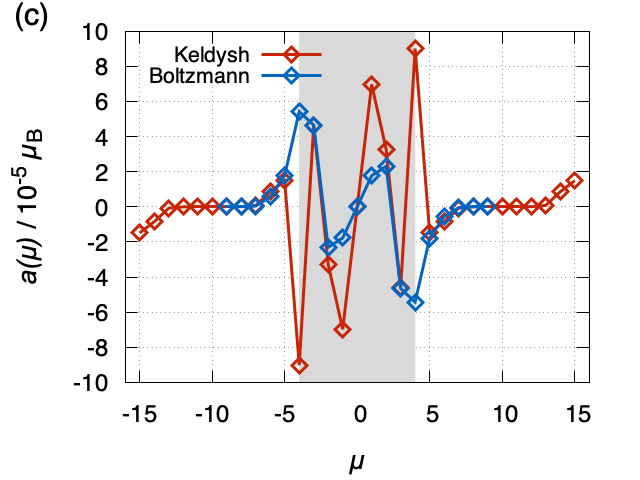}
	\caption{\label{fig:big} Magnetization for a thin film of 9 (a) Cu, (b) Pt and (c) U atoms (highlighted in grey). 
	The agreement is very good for Cu. U and Pt differ in absolute values. U also shows a more complex inner structure of the accumulation. While the extremal values of accumulation decrease for Cu and U, for Pt they increase. We attribute this to the strong influence of finite size effects.}
\end{figure}

\begin{figure}[htbp]
	\includegraphics[width=.47\textwidth]{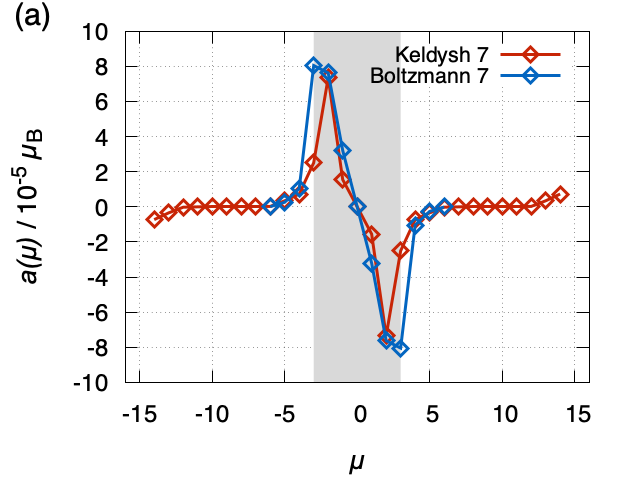}
	\includegraphics[width=.47\textwidth]{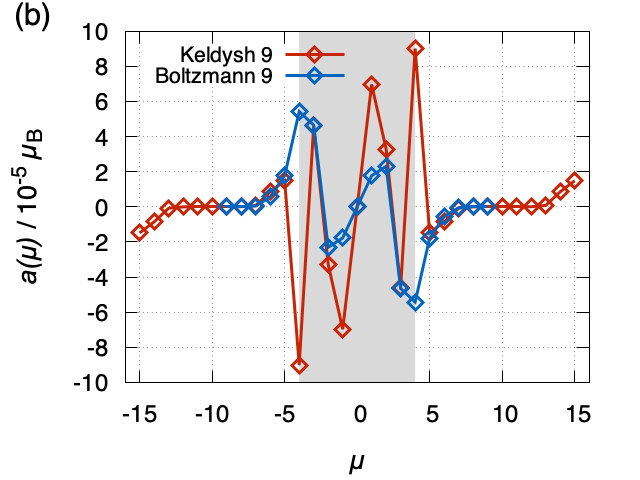}\\
	\vspace{.7cm}
	\includegraphics[width=.47\textwidth]{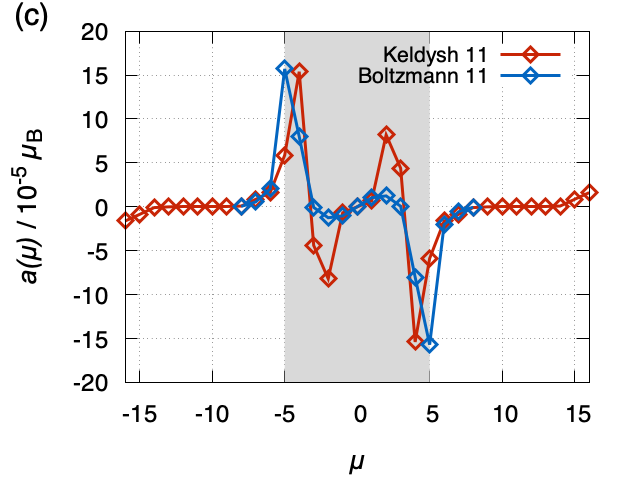}
	\caption{\label{fig:Uall} Magnetization for a thin film of (a) 7, (b) 9 and (c) 11 U atoms (highlighted in grey). 
	While the sign change is evident for the system with 9 layers, it vanishes for the smaller and the larger systems. The position of the extremal value in (a) and (c) differs slightly between the two methods.}
\end{figure}

\noindent The accumulations of the large systems (9 atomic layers) are depicted in Fig.~\ref{fig:big}. 
The behavior in (a) for Cu shows the best agreement between the two methods. The decay length of the accumulation is rather short compared to U and Pt. 
Pt in Fig.~\ref{fig:big} (b) shows a good agreement in extremal values, which may be a coincidence here, since for the small systems of Pt the extremal values differ by factor of roughly 2. The inner structure of the magnetization accumulation, differs more than that of Cu. 
For U in (c) with a more complex Fermi surface, the extremal values differ in sign. Additionally, U shows a more complex inner structure in the accumulation, which also differs between the two methods. 
We attribute the sign change which appears in the Keldysh calculation to finite size effects, since this deviation does not occur for a thinner system with 7 atomic U layers or a larger system of 11 atomic U layers. This is depicted in Fig.~\ref{fig:Uall} (a) and (c). For comparison, the system with 9 atomic U layers is depicted in Fig.~\ref{fig:Uall} (b) as well. 
The absolute extremal values match rather well, but especially for the 9 layer system and the 11 layer system the inner structure does differ significantly. While the extremal value of accumulation of Cu tends to decrease compared to the thin film with 3 layers, the extremal values for Pt increase in both methods. In the case of U, compared to the very thin film, each system shows a smaller extremal value. However, this extremal value is increasing from the 7 layer system to the 11 layer system except only for the 9 layer Boltzmann calculation. Also, the position of the extremal value slightly changes compared between Keldysh and Boltzmann. Again, we attribute this also to the strong influence of finite size effects in very thin films. 
However, the trend between the three systems remains qualitatively the same. For a detailed comparison of the smaller systems refer to the main text and Table~I in the main text.

\clearpage
\section{Fermi surfaces}
\begin{figure*}[hbtp]
\includegraphics[width=.25\textwidth]{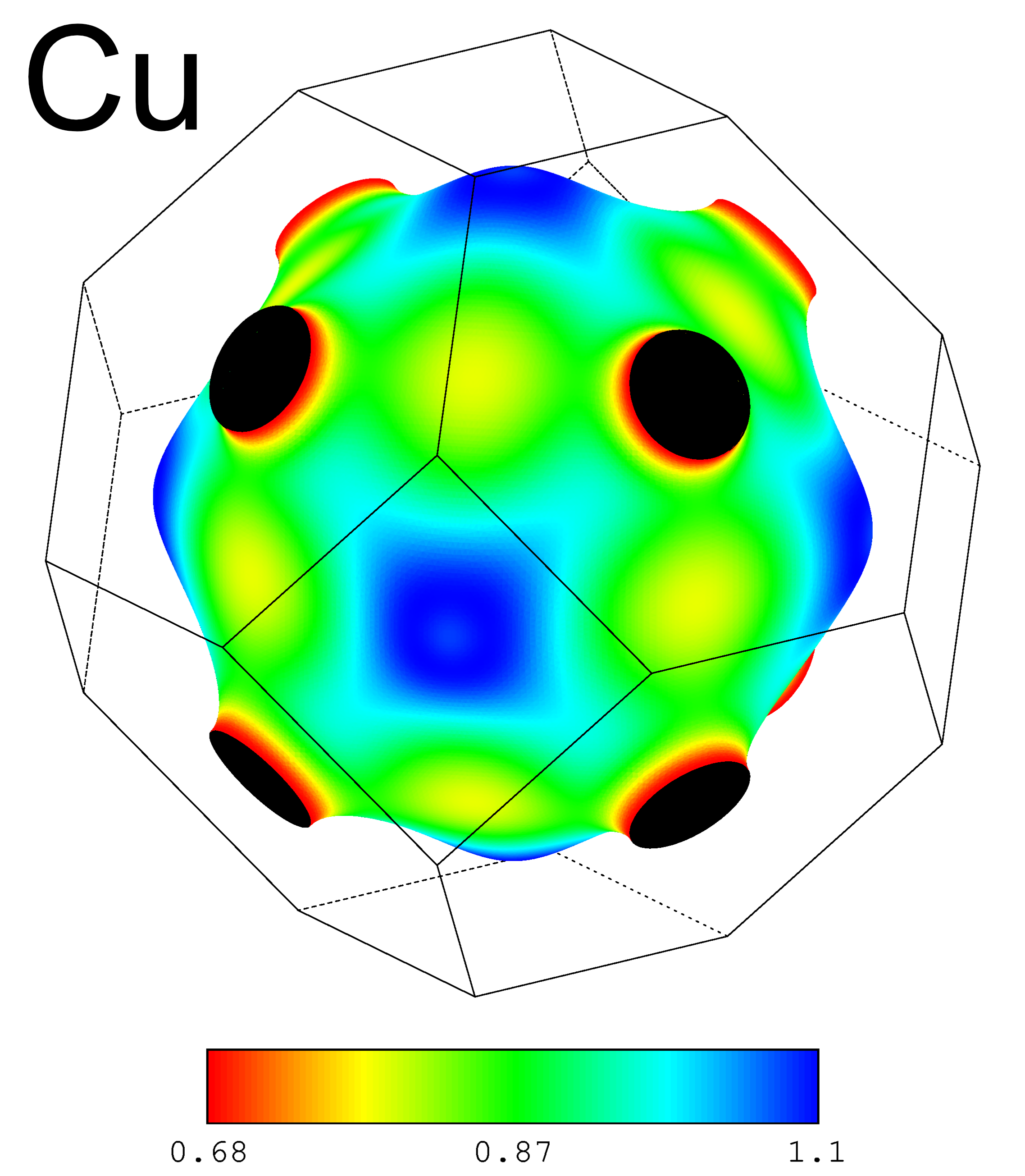}\hfill
\includegraphics[width=.25\textwidth]{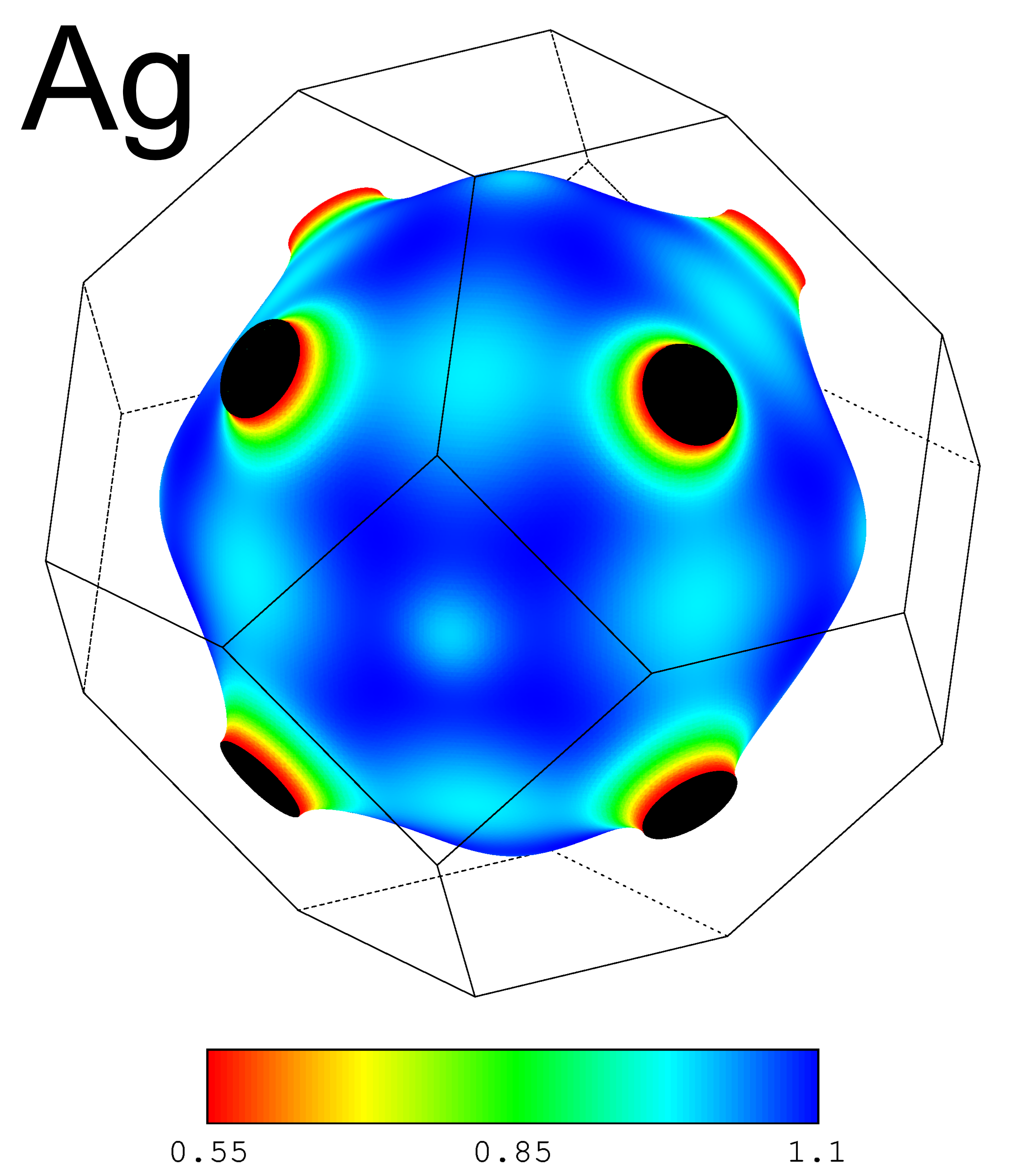}\hfill
\includegraphics[width=.25\textwidth]{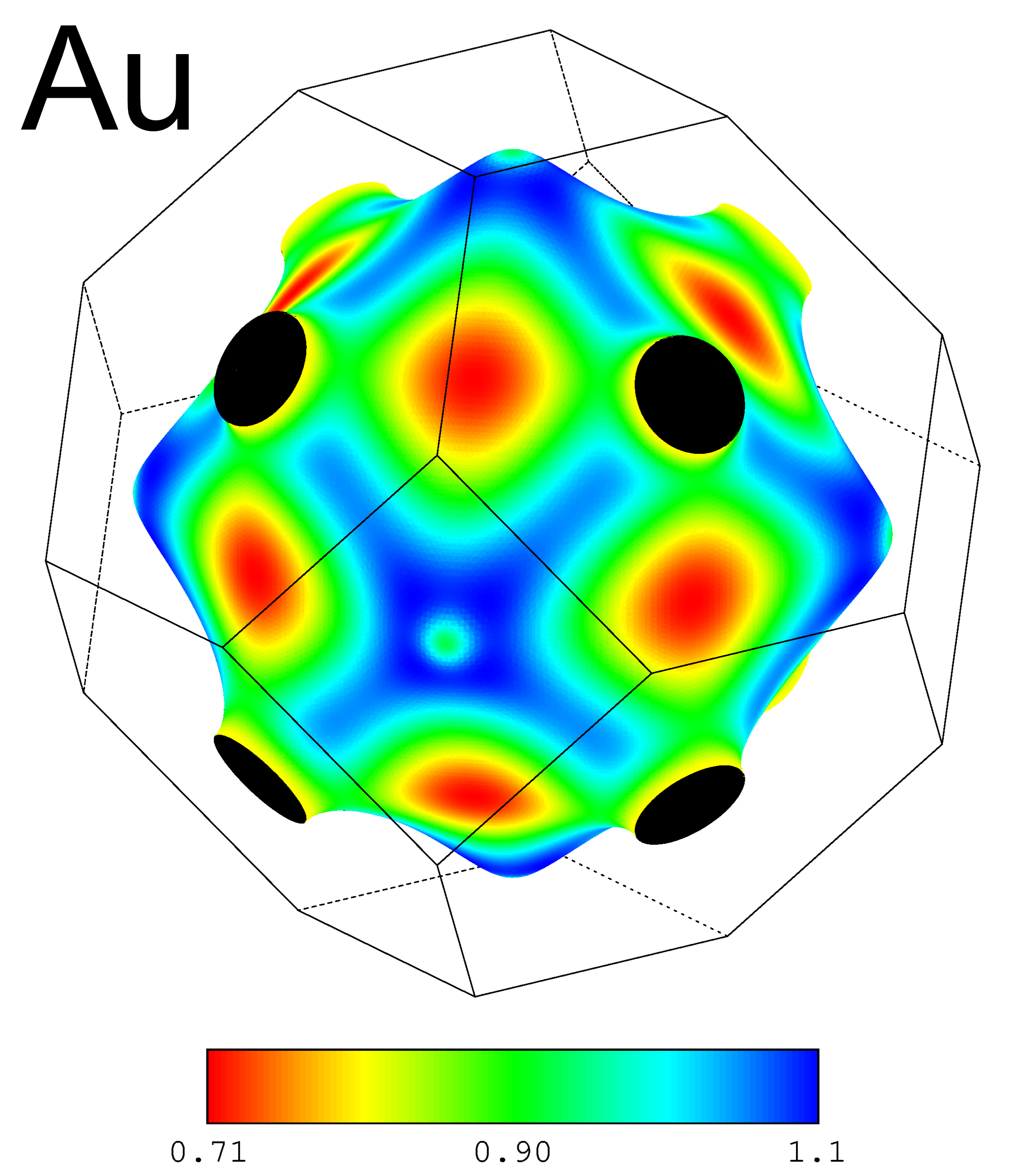}\\
\vspace{.8cm}
\includegraphics[width=.25\textwidth]{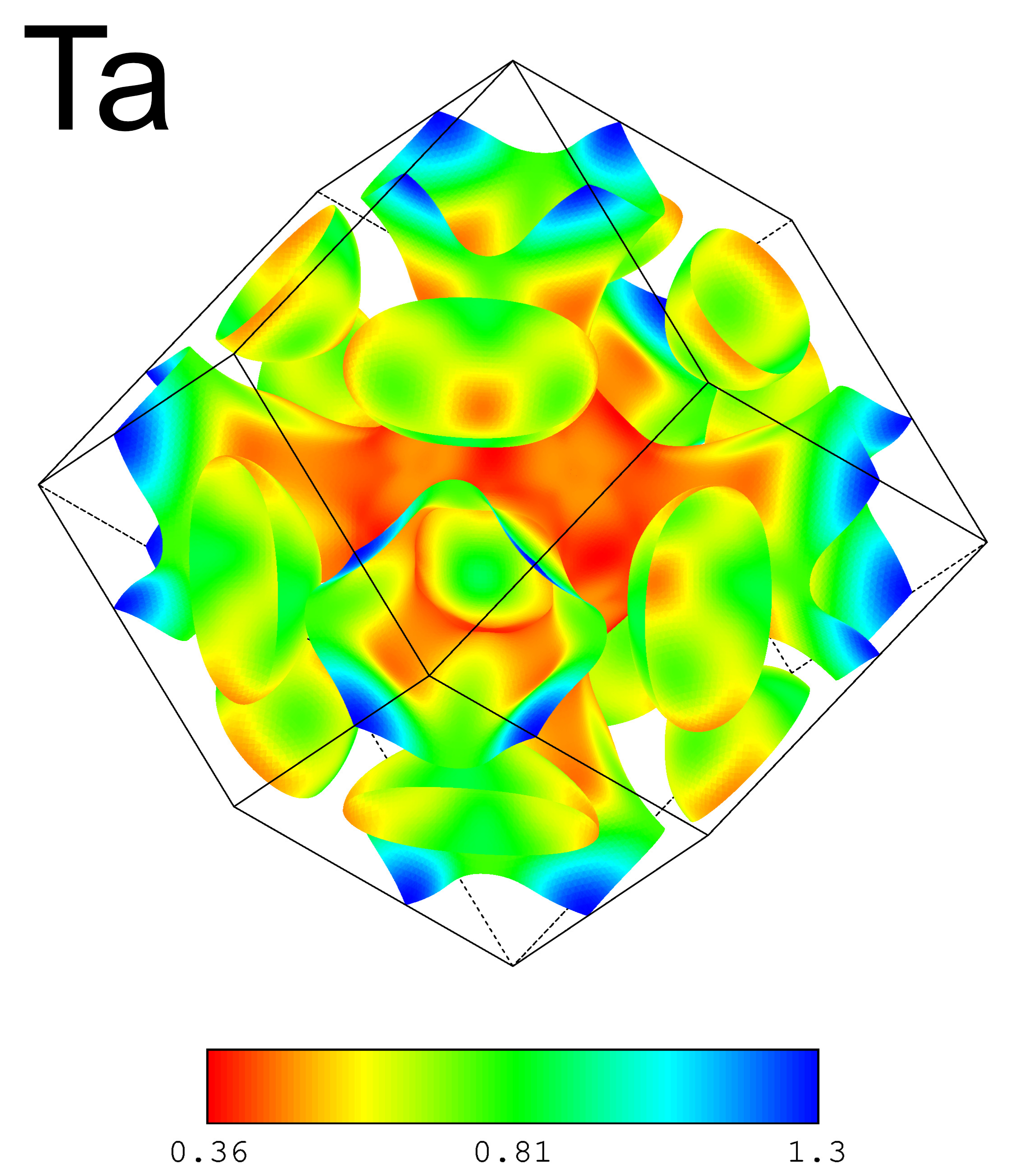}\hfill
\includegraphics[width=.25\textwidth]{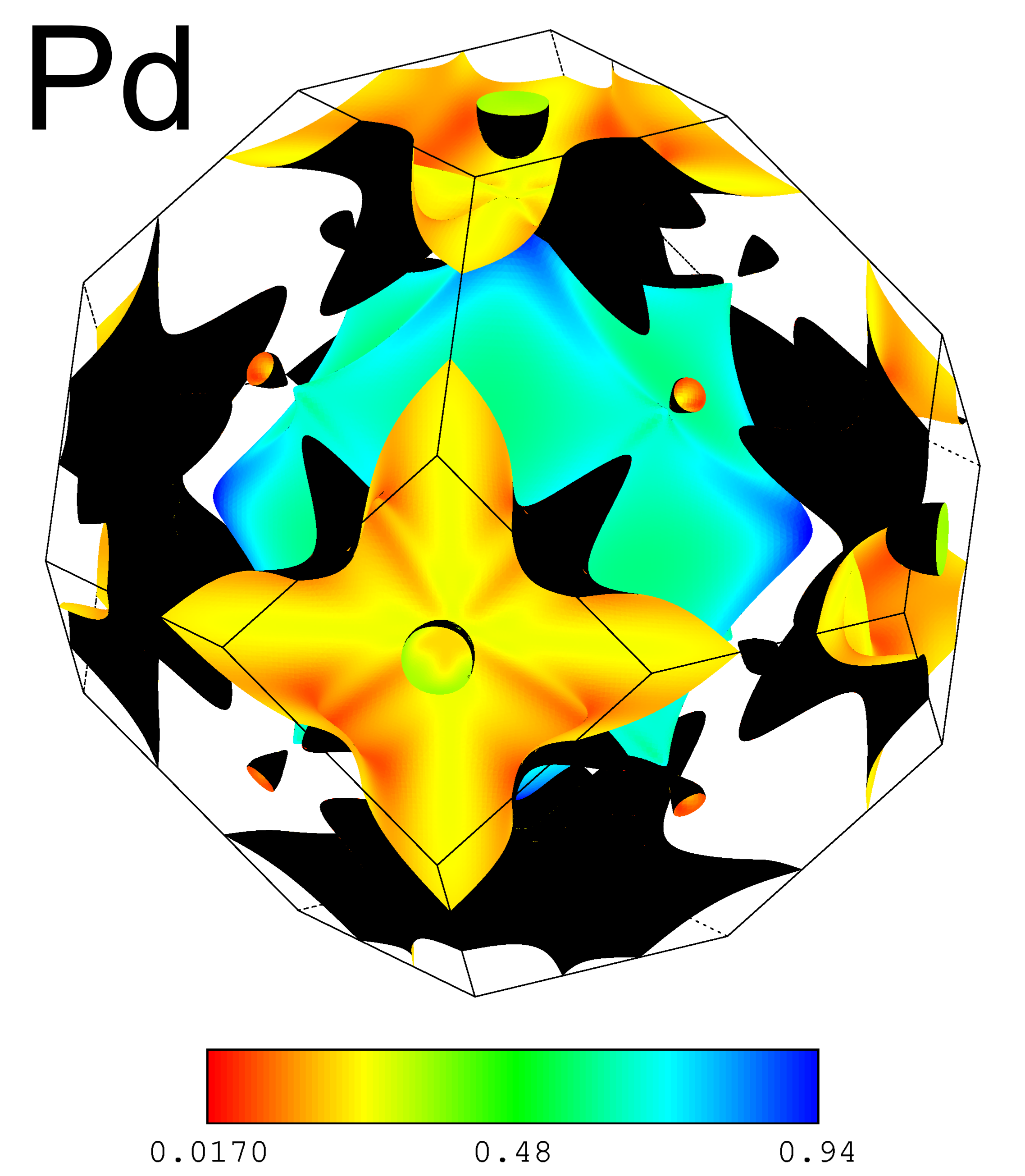}\hfill
\includegraphics[width=.25\textwidth]{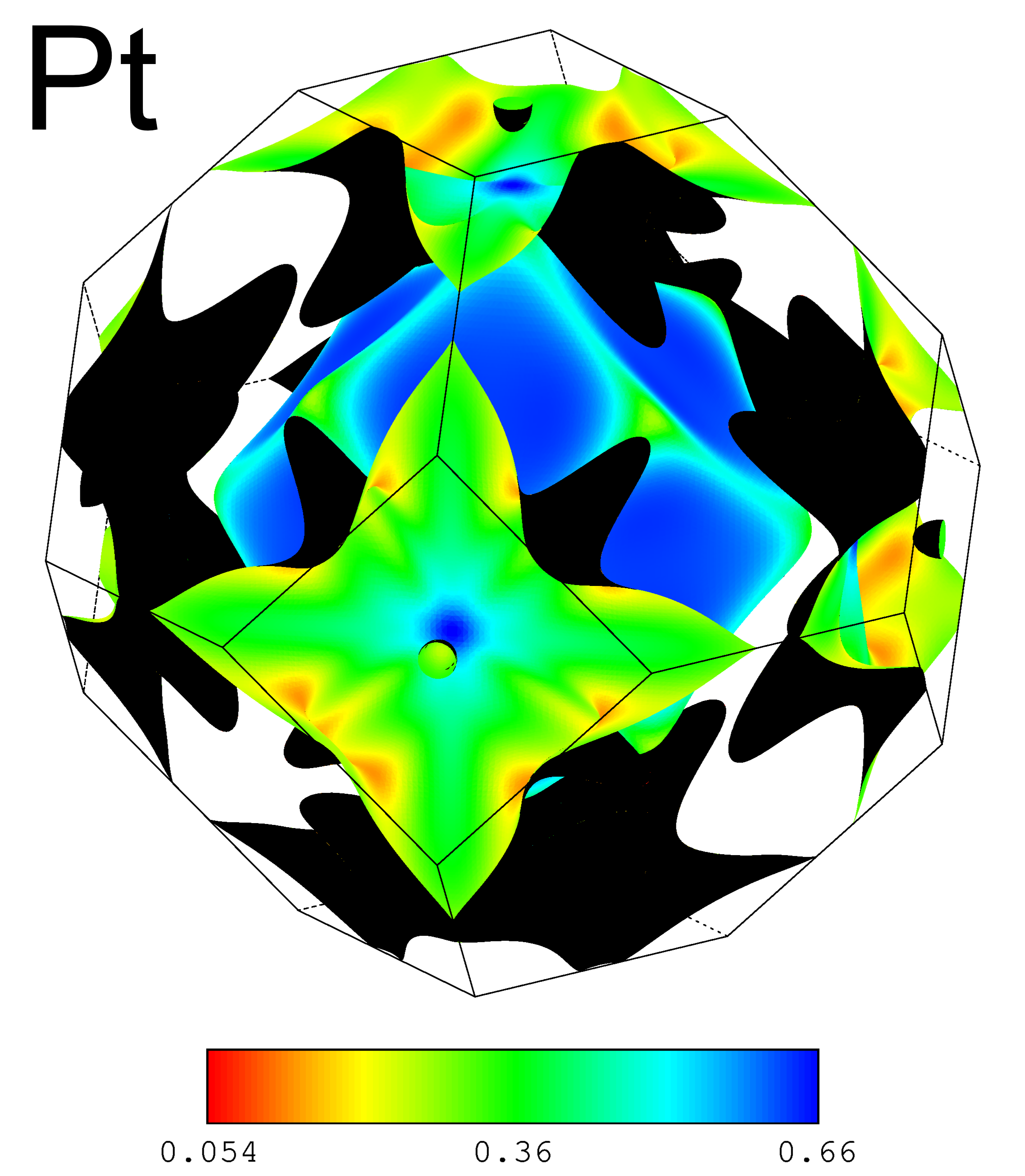}\\
\vspace{.8cm}
\includegraphics[width=.25\textwidth]{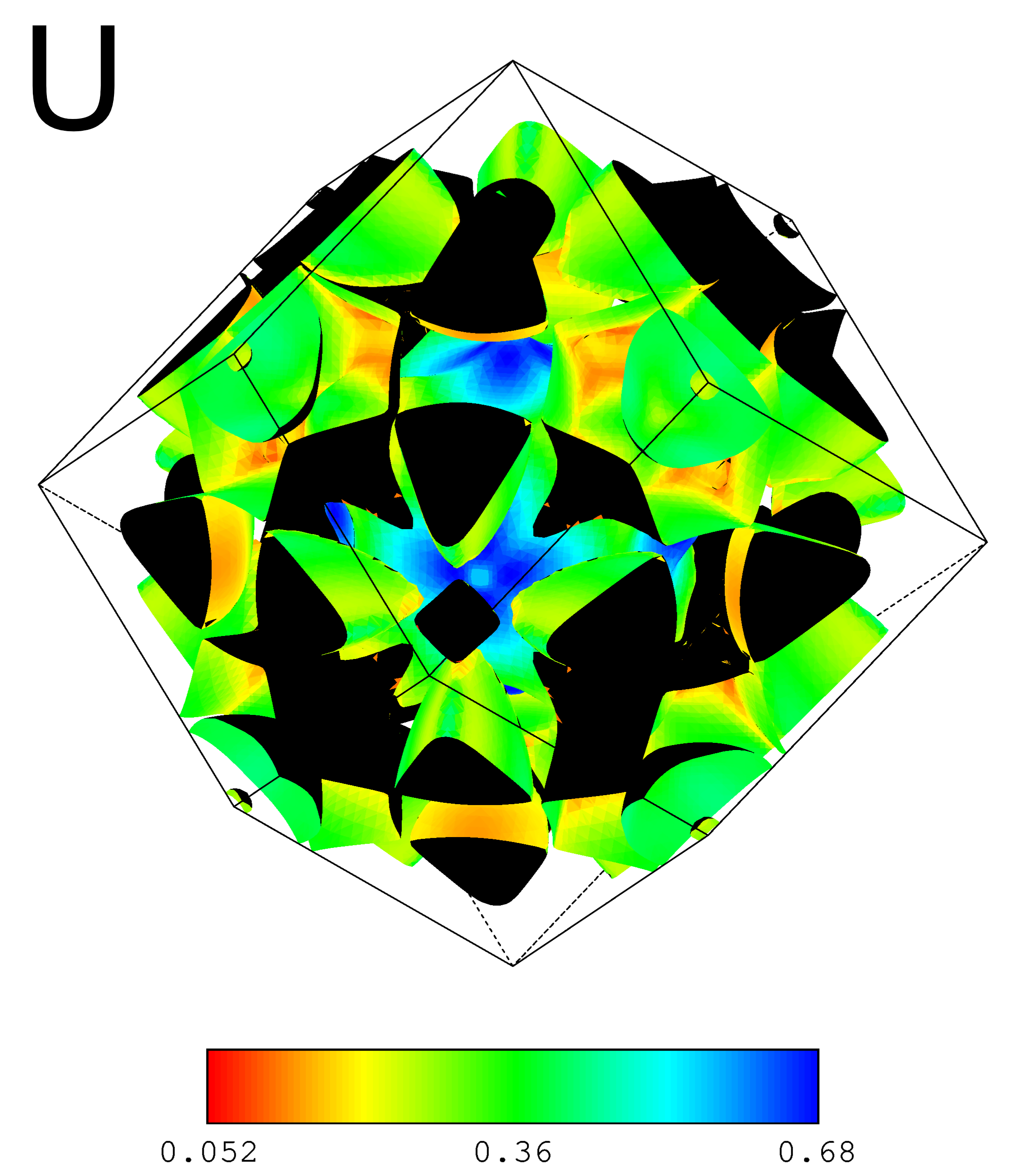}
\caption{\label{fig:fermi} Fermi surfaces of the all the different systems (Cu, Ag, Au, Ta, Pd, Pt, U). The color code depicts the group velocity on the Fermi surface.}
\end{figure*}

In Fig.~\ref{fig:fermi} the group velocity is colour encoded on the Fermi surface. 
Cu, Ag, and Au show simple Fermi surfaces without too much variation of the group velocity. Ta, Pd, Pt, and U show a more complex structure with a more elaborated variation of the group velocity on the surface.
For the entire discussion of the influences on the accumulation refer to the main text.

